\documentclass[10pt,prd,aps,amssymb]{article}
\usepackage{epsfig}

\usepackage{float}
\usepackage{graphicx}
\usepackage{cite} 
\usepackage[dvipsnames]{xcolor}
\usepackage{appendix}

\usepackage{tabu}
\usepackage{xcolor}
\usepackage{amssymb}

\begin{document}
\bigskip

\newcommand{\be}{\begin{equation}}
\newcommand{\en}{\end{equation}}
\newcommand{\bea}{\begin{eqnarray}}
\newcommand{\ena}{\end{eqnarray}}
\newcommand{\noi}{\noindent}
\newcommand{\ra}{\rightarrow}
\newcommand{\bib}{\bibitem}
\newcommand{\bff}{\begin{figure}}
\newcommand{\eff}{\end{figure}}
\newcommand{\refb}[1]{(\ref{#1})}

\begin{center}

{\Large \bf Thermodynamics and the Joule-Thomson expansion of dilaton black holes in 2 +1 dimensions}
\end{center}
\hspace{0.4 cm}

\begin{center}
Leonardo Balart\footnote{leonardo.balart@ufrontera.cl}\\
{\small \it Departamento de Ciencias F\'{\i}sicas, \\
Facultad de Ingenier\'{\i}a y Ciencias \\ \small Universidad de La Frontera, Casilla 54-D \\
Temuco, Chile.}\\
\vspace{0.3 cm}
Sharmanthie Fernando\footnote{fernando@nku.edu}\\
{\small\it Department of Physics, Geology  \& Engineering Technology}\\
{\small\it Northern Kentucky University}\\
{\small\it Highland Heights, Kentucky 41099, U.S.A.}\\
\end{center}

\begin{center}
{\bf Abstract}
\end{center}

In this paper, we study  thermodynamics and its applications of a family of  static charged dilaton black holes in 2+1 dimensions found by Chan and Mann \cite{Chan:1994qa} and  Xu \cite{Xu:2019pap}. There is a dimensionless parameter $N$ in the black hole solutions presented: it is related to the coupling constant for the dilaton with the electromagnetic field and the gravitational field. Black hole horizons exist only for  $ \frac{2}{3} \leq N < 2$. $N =1$ black hole is a solution to low energy string theory. Thermodynamics is studied in the canonical ensemble where charge is constant as well as in grand canonical ensemble where the potential is constant. The cosmological constant is considered as a thermodynamical variable where the pressure $P = -\frac{\Lambda}{ 8 \pi}$. We computed the first law for the black hole and introduced new  thermodynamical parameter in order to satisfy the first law. We computed  temperature, thermodynamic volume, specific heat capacities, Gibbs free energy  and studied local and global stability of the black hole. Thermodynamic volume differs from the geometric volume. In the canonical ensemble, we noticed that thermodynamic behavior falls into two broad categories: For $\frac{2}{3} \leq N < 1$, small black holes are locally stable and large black holes are not. For $ 1 \leq N < 2$ the black hole  is locally and globally stable for all values of the horizon radius. In order to demonstrate the two broad categories, we have presented  $N =1, \frac{2}{3}$ and $N = \frac{6}{7}$ black holes in detail. There were no phase transitions for the above values of $N$. In the grand canonical ensemble, we noticed that there is a Hawking-Page phase transition for the black hole with $N=6/5$. We have also studied  the Joule-Thomson expansion  and the Reverse Isoperimetric Inequality of these black holes. We made the observation that  the charged dilaton black hole does not violate the Reverse Isoperimetric Inequality for certain values of the parameters of the theory. Finally, we have suggested future work.

\hspace{0.7cm}

{\it Key words}: static, dilaton, black hole, anti-de Sitter space, first law, Joule-Thomson expansion, Hawking-Page phase transition, van der Waals phase transition.


\section{Introduction}

Black hole thermodynamics has been a very active area of research for the past 50 years or so since the seminal work of Bekenstein and Hawking where the relation of entropy to the area at the horizon and temperature at the horizon were discovered \cite{beken,hawking1}. Most interestingly, the work of Hawking and Page on anti-de Sitter black holes, where a first order phase transition between the thermal anti-de Sitter space and the Schwarzschild anti-de Sitter black hole was demonstrated, inspired many works on thermodynamics of anti-de Sitter black holes \cite{hawking2}. Another interesting property of anti-de Sitter black holes is that they have shown to have van der Waals type phase transitions between small and large black holes. For example, the Reissner-Nordstr\"om black hole has van der Waals type phase transitions \cite{chamblin}. In this work, such phase transitions were obtained by considering the correspondence between $(Q, \Phi) \leftrightarrow (P,V)$. Thermodynamics of the Born-Infeld AdS black holes were studied in grand canonical ensemble by Fernando in \cite{fernando5}. Thermodynamics and phase transitions of a black hole in massive gravity in canonial ensemble was studied by Fernando in \cite{fernando6}. In all three references \cite{chamblin}\cite{fernando5}\cite{fernando6}, the pressure was not a part of the thermodynamical variables.

A different approach to thermodynamics was considered (which is called extended phase space) where the cosmological constant is considered as the thermodynamic pressure of the black hole. In order for the black hole to have positive pressure, the cosmological constant has to be negative, leading to AdS black holes. This approach has attracted much attention in research of black hole thermodynamics. In this case, the first law of thermodynamics for black holes has an extra $V dP$ term and the mass $(M)$ is considered as the enthalpy $(H)$ rather than the internal energy \cite{kastor,dolan}. Many of the AdS black holes considered in extended pase space have van der Waals type phase transitions between small and large black holes. Some black holes have shown to have reentrant phase transitions.

Due to lack of space, we will only mention few examples of thermodynamics of black holes in AdS space in the extended phase space here: P-V criticality of Reissner-Nordstr\"om  AdS black hole was presented in \cite{mann3}. PV criticality of black holes in massive gravity was explored by Fernando in \cite{fernando1}; phase transitions and thermodynamic volume of thermodynamics of rotating black holes and black rings were studied by Altamirano et al. \cite{mann1}. Exhaustive study of thermodynamics of Born-Infeld black holes was done in \cite{mann2}.
Kerr-AdS analogue of triple point and solid/liquid/gas phase transitions was studied in \cite{mann14}. Thermodynamic phase transition of Euler-Heisenberg-AdS black hole on free energy landscape was studied by Dai et al. in \cite{dai}. Critical phenomena and reentrant phase transition of asymptotically Reissner-Nordstr\"om black holes were studied by Momennia and Hendi in \cite{hendi12}. Thermodynamics of quantum corrected black holes were studied by Wang et al. in \cite{wang2}. Some other works along those lines are \cite{liu2,hendi,pou,li,mo,zhang,mo2,cai, mirza, xu2,fernando7}. The reader may read a nice review on thermodynamics of AdS black holes by Kubiznak et al. \cite{kub} to see other examples related to this subject. The current status of the thermodynamics of black holes in extended phase space and related topics are given in a recent review by Mann \cite{mann5}.

2+ 1 dimensional black holes provide a simpler setting than 3+1 dimensions to understand complex details of technically challenging topics on black hole physics. After the publication of the well known BTZ black hole ~\cite{Banados:1992wn}, 2+1 dimensional black hole studies have become quite popular. Extensions of BTZ black hole with electrical and magnetic charge were done in \cite{Martinez:1999qi}, \cite{Kiem:1996ks}, \cite{Koikawa:1997am}. Some other works related to extension of BTZ black hole with charge with various other fields are \cite{Cataldo:1999wr} \cite{Cataldo:2000we} \cite{Gurtug:2010dr}.

String theory is one of the leading candidates for a theory of quantum gravity. Low energy string effective actions contain number of scalar fields and dilaton field is one of them. There have been many studies of black hole solutions of Einstein-dilaton gravity coupled to various other fields such as electrodynamics. It has been found that the spacetime structure and thermodynamics changes when a dilaton field is present
\cite{gibbons2} \cite{garfinkle}\cite{Dehghani:2018tcw} \cite{dehghani12}. In 2+1 dimensions, there are number of papers studying dilaton black holes: we will mention few of them even though this may not be the complete list. They are, \cite{Chan:1994qa}\cite{Xu:2019pap}\cite{Sa:1996ty}\cite{Chan:1996rd}\cite{Chan:1995wj}\cite{Sa:1995vs}\cite{Fernando:1999bh}\cite{Maki:2010zza}\cite{Dehghani:2017zkm}\cite{Dehghani:2018lat}\cite{Yamazaki:2001ue}\cite{Dehghani:2019noj}\cite{Dehghani:2022ntd}\cite{Dehghani:2022aae}\cite{Dehghani:2018hpb}\cite{Dehghani:2019dab}\cite{Younesizadeh:2020zvw}\cite{Dehghani:2019xhm}.

The goal of the current paper is to study thermodynamics of charged dilaton black holes in 2+1 dimensions derived by Chan and Mann ~\cite{Chan:1994qa} and Xu \cite{Xu:2019pap} : here the dilaton field is coupled to the electric field and the cosmological constant via an exponential function. BTZ black hole is one of the solutions of the action considered. Also, for certain parameters of the theory, the corresponding action becomes the low energy string action after a conformal transformation. This family of black hole solutions has very interesting properties and have been studied widely. One of our goals is to compare how the thermodynamics differs from the uncharged BTZ black hole and the charged BTZ black hole to this dilaton black hole. The BTZ black hole was studied in the extended phase space in \cite{mann2} and \cite{mann4}.

We would like to note that thermodynamics of  variety of dilaton black holes in 2+1 dimensions were studied in \cite{Dehghani:2017zkm}\cite{Dehghani:2018lat}\cite{Dehghani:2019noj}\cite{Dehghani:2022ntd}\cite{Dehghani:2022aae}\cite{Dehghani:2018hpb}\cite{Dehghani:2019dab}\cite{Dehghani:2019xhm} in the non-extended phase space.  In this paper, we do a thorough analysis of the dilaton black hole in the extended phase space. Also, we analyze the  local stability as well as the global thermodynamic stability by studying specific heat capacity and the  Gibbs free energy. 

In addition, as an application of thermodynamics, we also study the Joule-Thomson expansion for the dilaton field. As far as the authors knowledge, this is the only study of the Joule-Thomson expansion of a dilaton black hole in 2+1 dimensions. Joule-Thomson expansion of charged AdS black holes were studied by \"Okc\"u and Ayd\i{}ne \cite{okcu}. Joule-Thomson expansion of lower-dimensional black holes were studied by Liang et al. \cite{wang}. Holographic Joule-Thomson expansion in lower dimension was studied by Zarepour \cite{zarepour}. Joule-Thomson expansion of RN-AdS black hole immersed in perfect fluid dark matter was studied by Cao et al. \cite{cao1}. There are many works on studying Joule-Thomson expansion of black holes in literature. Few other references on Joule-Thomson expansion are \cite{wang3} \cite{morais2}. In addition we also have studied the Reverse Isoperimetric Inequality of the dilaton black hole.

The paper is organized as follows: in section 2, the charged dilaton  black hole in 2+1 dimensions is  introduced. Thermodynamics  for $\frac{2}{3} < N <2$  is presented in section 3. In section 4, thermodynamics for   $N = \frac{6}{7}$ is presented. In section 5, thermodynamics for  $N = 1$ is presented. In section 6, thermodynamics for   $N = \frac{2}{3}$ is presented. Comparison of thermodynamics of the dilaton black hole with the charged and neutral BTZ black hole in the canonical ensemble is given in section 7. Comments of thermodynamics and phase transitions in the grand canonical ensemble is given in section 8. Joule Thomson expansion is discussed in section 9. In section 10, Reverse Isoperimetric Inequality is discussed. Finally, the conclusion is given in section 11. There is an Appendix A which studies the thermodynamics of  neutral black holes and Appendix B where  the data for all the figures are given.


\section{Introduction to static charged dilaton black hole in 2+1 dimensions}
We consider the Einstein-Maxwell-dilaton action for 2+1-dimensions given by Chan and Mann in Ref.~\cite{Chan:1994qa}
\begin{equation}
S = \int d^3x \sqrt{-g} \, \left(R - \frac{B}{2}(\nabla\phi)^2 -  e^{-4 a \phi} F_{\mu\nu}F^{\mu\nu} - V(\phi) \right)   \, ,
\label{action}
\end{equation}
where $R$ is the Ricci scalar, the third term in the parentheses represents the coupling between electrodynamics and the dilaton field $\phi$, $F_{\mu\nu}$ is the Maxwell field strength, the dilaton potential is $V(\phi) = 2 e^{b \phi}\Lambda$, $\Lambda$ is the cosmological constant and $a$, $b$ and $B$ are coupling constant. Here, when $\Lambda <0$ corresponds to anti-de Sitter space and $\Lambda >0$ corresponds to the de Sitter space. Notice that in the paper by Chan and Mann \cite{Chan:1994qa}, $\Lambda >0$ is treated as the anti-de Sitter case and $\Lambda <0$ is considered as the de Sitter case. To align our work with most literature on anti-de Sitter black holes, we will consider $\Lambda <0$. Hence, when $\Lambda$ appears in our work, it will be a $-\Lambda$ in Chan and Manns work~\cite{Chan:1994qa}.

First, we will present the black hole solutions derived by Chan and Mann in \cite{Chan:1994qa}. They considered the following line element,

\begin{equation} \label{metric1}
ds^2 = - f(r) dt^2 + f^{-1} (r) dr^2 +  r^2 R^2(r) d\theta^2 \, ,
\label{element}
\end{equation}
where 
\begin{equation} \label{metric2}
R(r) = \gamma \, r^{\frac{N}{2}-1}
\label{R-def}
\end{equation}
and
\begin{equation} \label{metric3}
 f(r) = - \frac{2 m r^{1-\frac{N}{2}}}{N} -\frac{ 8 \Lambda \beta^{2-N}r^N}{(3N-2)N}  + \frac{ 8 Q^2}{(2-N) N}
  \, .
\label{metric-f}
\end{equation}
Here $\gamma$ is a integration constant. Note that the dimensions are: $[r] = L$, $[m] = L^{\frac{N}{2}-1}$, $[\Lambda]= L^{-2}$, $[\beta]= L$, $[Q]= L^{0}$ and $[\gamma]=  L^{1-\frac{N}{2}}$, where $L$ represents units of length. 
The dilaton field is given by,
\begin{equation} \label{phi}
\phi = k \ln\left( \frac{r}{\beta}  \right)  \, ,
\label{}
\end{equation}
where,
\begin{equation} \label{kvalue}
k = \pm  \sqrt{\frac{N(2-N)}{2B}};
  \hspace{0.4 cm}  4 a k = b k = N - 2
\label{}
\end{equation}
The term $\beta$ represents a constant of integration. Later in the paper $\beta$ will be presented as a   thermodynamic quantity. We will  choose $\beta >0$ since it appears in the $\ln(r/\beta)$ for $\phi$ in eq.$\refb{phi}$. 
The dimensions of the parameter $\gamma$ are to keep the angular part of the metric dimensionless. However, it can be absorbed by redefining the angular coordinate, so it does not introduce physical scale and therefore cannot be considered a thermodynamic variable.

Notice that $N$ is a dimensionless parameter and is related to the coupling constant for the dilaton with the electromagnetic and gravitational field. Varying $N$ leads to  a family of black hole solutions. From eq.$\refb{kvalue}$, it is clear that $ 2> N>0$. As discussed in detail by Chan and Mann \cite{Chan:1994qa}, black hole event horizon exists only for  $2>N>2/3$ and $\Lambda <0$ for the metric in eq.$\refb{metric3}$. There is no general method to compute black hole horizons for a general $N$ value. The authors in \cite{Chan:1994qa}, discussed black holes with  $ N = \frac{6}{5}, \frac{6}{7}, \frac{4}{3}, \frac{4}{5}$.

It is important to make comments on how above black hole solutions are related to string theory: notice that the low energy action for string theory in 2+1 dimensions (setting some of the gauge fields and fermions  to be zero) is given by \cite{Chan:1994qa} \cite{maki},
\begin{equation}
S_{String} = \int d^3x \sqrt{-g^S} \, e^{-2\phi} \left(R[g^S] + 4 (\nabla\phi)^2 -  F_{\mu\nu}F^{\mu\nu} - 2 \Lambda \right)   \, ,
\label{actionstring}
\end{equation}
Here, $g^S$ is the metric in the string frame. One can obtain the string action in eq.$\refb{actionstring}$  from the Einstein-Maxwell-dilaton action given in eq.$\refb{action}$ as follows:
If  the parameters in the action eq.$\refb{action}$ are chosen to be $b = 4 a = 4 = B/2$, the string action and Einstein-Maxwell-dilaton action will be related  by the following conformal transformation.
\begin{equation}
g_{\alpha \beta}^{S} = e^{2 \phi} g_{\alpha \beta}^{Einstein}
\end{equation}
It is clear that the above parameters will yield $N = 1$ and $ k = -1/4$. When $N=1$ is substituted to  $f(r)$ in eq.$\refb{metric3}$ and $R(r)$ in eq.$\refb{metric2}$, the metric for the charged $N=1$ black hole is obtained as
\begin{equation} \label{metricn1c}
ds^2 = - ( - 8 \Lambda \beta r - 2 m \sqrt{r} + 8 Q^2) dt^2 + \frac{ dr^2}{(- 8 \Lambda \beta r - 2 m \sqrt{r} + 8 Q^2)} +  \gamma^2 r d\theta^2 \, ,
\label{metricn1c}
\end{equation}  
There is a (1+1) black hole solution to string theory known as MSW black hole \cite{Mandal:1991tz}. By taking the product of this black hole with $S^1$, another black hole is generated in string theory known as 2+1 MSW black hole which has no charge. In this case, when the 2+1 MSW black hole is brought to Einstein frame by a conformal transformation, it becomes the uncharged dilaton black hole for N=1 case given by,
\begin{equation} \label{metricn1u}
ds^2 = - ( - 8 \Lambda \beta r - 2 m \sqrt{r}) dt^2 + \frac{ dr^2}{(- 8 \Lambda \beta r - 2 m \sqrt{r})} +  \gamma^2 r d\theta^2 \, ,
\label{elementn1u}
\end{equation} 
Now, solutions in string frame enjoy duality properties such that one can take an uncharged solution and perform duality transformation to obtain charged solutions: a detailed explanation on this topic is given  with examples by Horowitz in \cite{Horowitz:1992jp}. These dualities are in essence due to symmetries in the string theory and has been utilized to generate new solutions \cite{sen1}\cite{sen2}. 
If a duality is performed to the neutral 2+1 MSW black hole in string frame and is brought to the Einstein frame by a conformal transformation, then the charged black hole for N=1 given by eq.$\refb{metricn1c}$ can be obtained. Hence N=1 charged black hole is dual to the uncharged N=1 black hole and is known as 2+1 charged MSW black hole. Hence studying thermodynamics of the solution for $N=1$ is important. We will compute thermodynamic quantities for the N=1 case in the Einstein frame.

One can observe in eq.$\refb{metric3}$, there is a singularity at $ N = 2/3$ in $f(r)$. This case was not discussed in Chan and Mann's paper. However, Xu \cite{Xu:2019pap}, derived black hole solutions for $N = \frac{2}{3}$ to be the following metric:

\begin{equation} 
f(r) =  -\frac{24 M r^{2/3}}{\gamma} - 6 \beta^{4/3} \Lambda \, r^{2/3} \log \left(\frac{r}{\beta }\right)  + 9 Q^2  \, .
\label{}
\end{equation}
Since there are two  metric functions depending on the value of $N$, we will discuss thermodynamics separately for $ \frac{2}{3} < N < 2$ and for $N = \frac{2}{3}$.


\section{General expression for thermodynamics of the black hole for  
$2/3< N <2$}
\label{S-F-N}
In this paper we are studying thermodynamics in the extended phase space where we consider pressure  $ P = -\frac{\Lambda}{ 8 \pi}$. In this section we will address thermodynamics of the dilaton black hole for  general values of $N$ in the range $2/3 < N < 2$ given by the metric in eq.$\refb{metric1}$, eq.$\refb{metric2}$ and eq.$\refb{metric3}$. We will discuss the special case for $N = 2/3$ n section 6.

First, we will replace the mass $m$ with mass $M$ in $f(r)$ with the following relations as given in \cite{Dehghani:2019xhm},
\begin{equation} \label{mvalue}
M = \frac{\gamma \, m}{8}    \, 
\label{mass-16-b}
\end{equation}
Then, $f(r)$ is given by,
\begin{equation} \label{metric4}
 f(r) = - \frac{16 M r^{1-\frac{N}{2}}}{N \gamma} -\frac{ 8 \Lambda \beta^{2-N}r^N}{(3N-2)N}  + \frac{ 8 Q^2}{(2-N) N} \,.
\label{metric-f-N}
\end{equation}
Considering the fact that $f(r_+) = 0$  we get
\begin{equation}
M = \frac{\gamma}{2} r_+^{\frac{N}{2}-1}   \left( - \frac{\Lambda\beta^{2-N}  r_+^N }{3 N - 2}-\frac{Q^2}{N-2}\right)    \, .
\label{mass-M-N}
\end{equation}

\subsection{First law of black hole thermodynamics}
\label{First-law-N}
Notice that we consider the mass $M$ as the enthalpy $H$ in the rest of the work. We can obtain the entropy using the area law as,
\begin{equation}
S = \frac{2 \pi r_+ R(r_+)}{4} = \frac{\pi}{2} \gamma \, r_+^{N/2}    \, .
\label{entropy-S-N}
\end{equation}
It is possible to rewrite the mass $M$ in terms of $S,P,Q, \beta, \gamma$ as,
\begin{equation}
M(S,P,Q,\beta, \gamma) = 
\frac{\pi ^{\frac{2}{N}-2} \beta^{-N} S^{-2/N} \gamma^{\frac{2}{N} - 2}\left[32 \beta^2 (N-2) P S^3-\pi  \gamma^2 (3N - 2) Q^2 S \beta^N\right]}{4^{1/N}  (N - 2) (3 N - 2)} \, .
\label{mass1}
\end{equation}
We can write the following expression for the first law of thermodynamics as,
\begin{equation}
dM = \left ( \frac{ \partial M} { \partial S } \right) _{P,Q,\beta}  dS + \left ( \frac{ \partial M} { \partial P } \right) _{S,Q,\beta} dP + \left ( \frac{ \partial M} { \partial Q } \right) _{S,P, \beta}  dQ
+ \left(\frac{\partial M}{\partial \beta}\right)_{S,P,Q} d\beta 
\,\,\label{dm1} \,  ,
\end{equation}
or 
\begin{equation}
dM = T dS + V dP + \Phi dQ + \Omega d\beta 
\,\,\label{} \,  .
\end{equation}
Then the temperature can be calculated from the first law of black hole thermodynamics as
\begin{equation} \label{tempN}
T = \left(\frac{\partial M}{\partial S}\right)_{P,Q,\beta}  = \frac{-\Lambda \beta ^{2-N}  r_+^N - Q^2}{N \pi  r_+}    \, 
\label{}
\end{equation}
The Hawking temperature obtained using the surface gravity is
\begin{equation}
 T_{H}   =  \frac{1}{4 \pi} \left|\frac{dg_{tt}}{dt} \sqrt{-g^{tt} g^{rr}}  \right|_{r_+}  =   \frac{2 M (N - 2) r_+^{-N/2}}{\pi N \gamma } - \frac{2 \Lambda \beta ^{2-N}  r_+^{N-1}}{\pi (3 N - 2)}
 = \frac{- \Lambda \beta ^{2-N}  r_+^N - Q^2}{N \pi  r_+} 
  \, ,
\label{tempn}
\end{equation}
which is equal to $T$. We also note that the black hole can have degenerate horizons where the temperature is zero; in this case, the degenerate horizon is given by, $r_{ex}$,
\be \label{rex1}
r_{ex(N)} = \left(\frac{ Q^2}{  - \Lambda \beta^{2 - N}} \right)^{1/N}
\en
One can obtain the corresponding thermodynamic volume as,
\begin{equation} 
V = \left(\frac{\partial M}{\partial P}\right)_{S,Q,\beta}  = \frac{4 \pi  \gamma \, \beta^{2 - N} r_+^{\frac{3 N}{2}-1}}{3 N - 2}  \, 
\label{volumen}
\end{equation}
which is different from the geometric volume, $\pi r_+^2$. Such behavior has been observed in charged BTZ black hole \cite{mann4} and non-linear black holes in 2+1 dimensions \cite{fernando8}.
The electric potential at the horizon is given by
\begin{equation} 
\Phi = \left(\frac{\partial M}{\partial Q}\right)_{S,P,\beta}  = \frac{\gamma \, Q }{(2 - N) \, r_+^{1-\frac{N}{2}} }   \, .
\label{}
\end{equation}
Finally, we get the thermodynamic quantity $\Omega$ conjugate to the parameter $\beta$
\begin{equation} 
\Omega = \left(\frac{\partial M}{\partial \beta}\right)_{S,P,Q} = \frac{(N - 2) \gamma \, \beta^{1-N} \Lambda \, r_+^{\frac{3 N}{2}-1}}{2 (3 N - 2)} \, .
\label{Omega-sta}
\end{equation}

We can note that the thermodynamic quantities have the following dimensions in terms of $L$, the unit of length:
$[T] = L^{-1}$, $[S] = L^{N/2}$, $[P] = L^{-2}$, $[V] = L^{2-N/2}$, $[Q] = L^{0}$, $[\Phi] = L^{0}$, $[\beta] = L$, and $[\Omega] = L^{-1}$. Later we will obtain $[G] = L^{0}$ and $[C_{P,Q}] = L^{N/2}$. All the thermodynamic quantities are considered as geometric quantities and one can make the relation to physical quantities with dimensional analysis along \cite{mann2}. For example, in D=3, [Press] = $\frac{ h c}{l_P} [P]$ and [Temp] = $\frac{hc}{\kappa} [T].$ In \cite{mann2}, specific volume $v$ of the black hole was identified with the horizon radius as $v = 4 r_+$ for $D=3$.

Here, the parameter $\beta$ acts as a dilaton scale parameter, fixing the reference point of the scalar field where $\phi = 0$ and setting the energy scale at which dilaton–matter coupling effects become relevant. In contrast to $\beta$, the parameter $\gamma$ cannot be considered a thermodynamic variable because it only represents a normalization of the angular coordinate; that is, it does not change the physical state of the black hole.

We have introduced $\beta$ as a thermodynamic variable. To justify this, we provide below a Euclidean derivation showing that the associated conjugate charge
$\Omega$ is well defined and integrable. For the present case, the Euclidean action in the canonical ensemble is given by \cite{Gibbons:1976ue}
\begin{equation}
I_E = \beta_E (M - TS - \Phi Q - \Omega \beta)
\,\,\label{Eu-action} \,  .
\end{equation}
where $\beta_E = T^{-1}$ is the Euclidean period, fixed by the regularity condition at the horizon. The Euclidean action at fixed pressure ($\Lambda$ constant). The conjugate charge $\Omega$ associated with the dilaton scale $\beta$, is then defined as the variation of the free energy $I_E/\beta_E$, evaluated on-shell, with respect this physical parameter
\begin{equation}
\Omega = - T \left( \frac{\partial I_E}{\partial \beta}\right)_{\mbox{on-shell}}
\,\,\label{Eu-Omega} \,  ,
\end{equation}
where the derivative is taken with all other thermodynamic variables kept fixed. This procedure yields an integrable form of the first law, and the resulting expression for $\Omega$ is in full agreement with the one derived in eq.$\refb{Omega-sta}$. In Addition, this definition satisfies the thermodynamic integrability condition
\begin{equation}
 \left( \frac{\partial T}{\partial \beta}\right) =  \left( \frac{\partial \Omega}{\partial S}\right) = \frac{(N-2) \Lambda  \beta ^{1-N} r_+^{N-1}}{N \pi  }
\,\,\label{} \,  .
\end{equation}


\subsection{Thermodynamical quantities}

We will calculate few more thermodynamical quantities  as follows: specific heat capacity is given by,
\begin{equation} \label{specificN}
C_{{P,Q}} =  T \left( \frac{\partial S}{\partial T} \right)_{P} =
\frac{N \pi  \gamma \,   r^{N/2}_+ \left( \beta^2 \Lambda r^N_+  + Q^2 \beta ^N\right)}{4 (N - 1)  \beta^2 \Lambda r^N_+ - 4 Q^2 \beta^N} 
\label{C-N} 
\,
\end{equation} 
Combining eq.$\refb{tempN}$ and $\Lambda = -8 \pi P$, and substituting for $r_+$ in terms of specific volume $v$, one can obtain the state equation as,
\be \label{state}
P = \frac{N T 4^{N-1}}{ 8 \beta^{2 - N} v^{N-1}} + \frac{ Q^2 4^N}{ 8 \pi  v^N \beta^{2 - N}}  \, .
\en

In order to understand the global stability  of the black holes thermodynamically, one has to study the Gibbs free energy, G. G is given by $H- T S$, where $H$ is the enthalpy. In the extended phase space enthalpy $H = M$. Hence,  $G = M - TS$. Hence, the Gibbs free energy is given by the following expression:
\begin{equation} \label{gibbsN}
G = \frac{\gamma \, r_+^{\frac{N}{2}-1} }{N} \left(\frac{ (N-1) \beta^{2-N} \Lambda  r_+^N}{3 N - 2}-\frac{Q^2}{N - 2}\right)
\,\,\label{gibbs-N} \,  
\end{equation} 

\subsection{Comments on the maximum temperature $T_{max}$ and $C_{P,Q}$}

\begin{figure} [H]
\begin{center}
\includegraphics[scale=0.80]{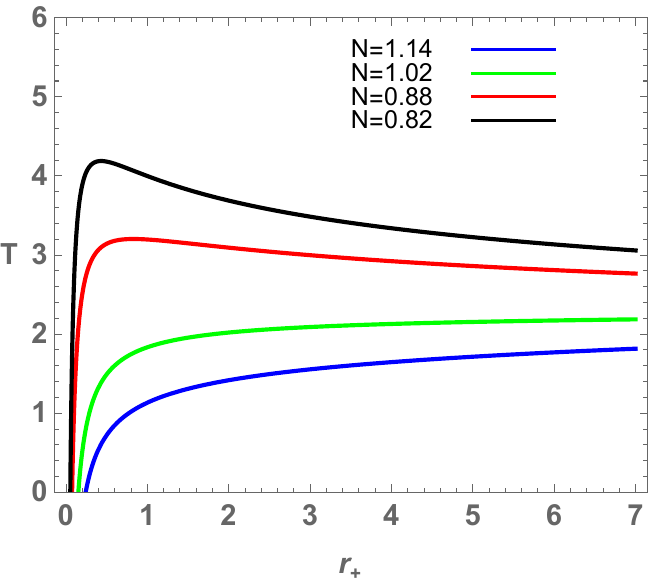}  

\caption{The figure shows   $T$  vs  $r_+$ for varying values of $N$. Here $Q = 1,\beta = 12.88, \Lambda = -0.562$.
}
\label{tvsn}
 \end{center}
 \end{figure}

\begin{figure} [H]
\begin{center}
\includegraphics[scale=0.80]{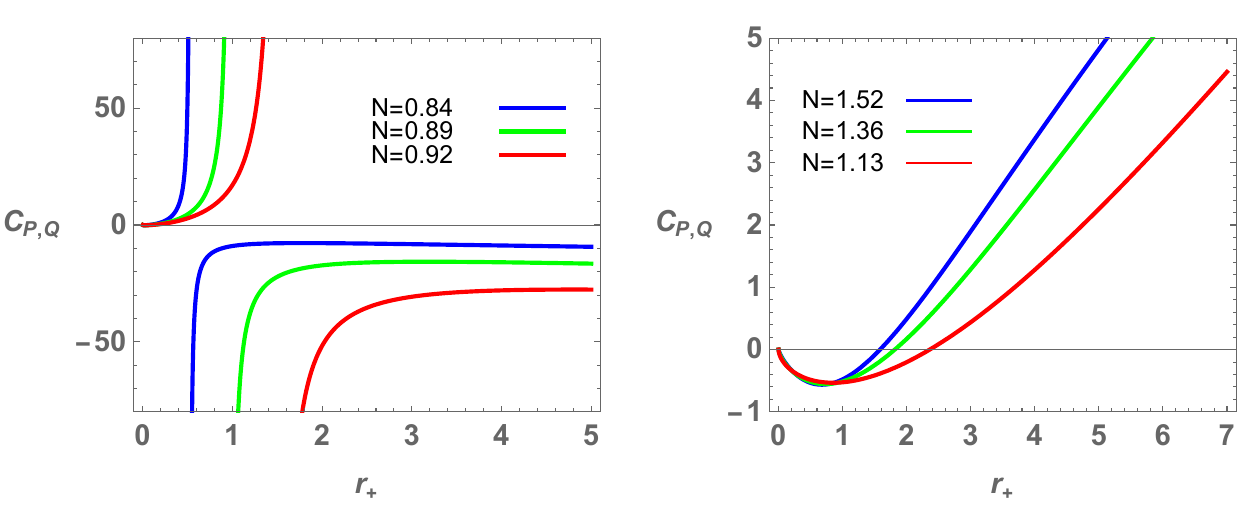}  

\caption{The figure shows   $C_{P,Q}$  vs  $r_+$ for varying values of $N$. Left graph is for $N < 1$ and the right graph is for $N > 1$. Here $ \Lambda = -0.1, \gamma = 1$ for both graphs. For the left graph, $Q = 0.5, \beta = 17$ and for the right graph $Q = 0.386, \beta = 0.5$
}
\label{cpalln}
 \end{center}
 \end{figure}

Before proceeding to discuss more details on thermodynamics, we would like to make comments on the relation of the temperature and value $N$ as follows:
The black hole has a maximum temperature $T_{max}$ at,
\begin{equation}
r_{max} = \frac{Q^2}{ ( N-1) \beta^{2 - N} \Lambda}
\end{equation}
Since $\Lambda <0$, $r_{max}$ will be real only if $N < 1$. Hence for $ 2/3 < N < 1$, the maximum temperature  is given by,
\be
T_{max} = \frac{Q^2}{ \pi (1 - N) r_+}
\en
When $ N \geq 1$, the temperature does not have a maximum. In Fig.$\refb{tvsn}$, temperature is plotted for various values of $N$ and one can see that for $N <1$ there are maximum values of the temperature.
Existence of a maximum for the temperature does have an effect on the $C_{P,Q}$. Since $C_{P,Q} = T \left( \frac{\partial S}{\partial T} \right)_{P}$ which is also equal to 
$T \left( \frac{\partial S}{\partial r_+} \right)_{P} / \left( \frac{\partial T}{\partial r_+} \right)_{P}$, existence of the maximum of the temperature does significantly effects the behavior of $C_{P,Q}$. Since $\left( \frac{\partial T}{\partial r_+} \right)=0$ at $T_{max}$, $C_{P,Q} \ra \pm \infty$ at $r = r_{max}$. Hence there would be two branches for $C_{P,Q}$ when $T_{max}$ exists where as there will be only one branch when $T_{max}$ does not exist. This phenomenon is demonstrated in Fig.$\refb{cpalln}$.

We will discuss  the details of $C_{P,Q}$ in more depth and detail for the following two examples: we have chosen to describe thermodynamic stability of a black hole with $N=6/7$ and one with $N = 1$ which represents black holes with a maximum temperature and one without.

\section{Thermodynamics for the black hole with $N= \frac{6}{7}$}

In the previous section, it was clear that thermodynamics will be different for black holes with $2/3 < N < 1$ and for $ 1 \leq N < 2$. Hence, in order to 
make quantitative analysis of  thermodynamics of black holes, we will analyze black holes with  $N= \frac{6}{7}$ (which falls in the region $2/3 < N < 1$) and $ N = 1$ (which falls in the region for $ 1 \leq N < 2$). In section 6, we will separately address $N = 2/3$ case since it has a seperate metric. Studying $N=1$ case is noteworthy since it is related to string theory.


\subsection{Horizon structure}

\begin{figure} [H]
\begin{center}
\includegraphics[scale=0.80]{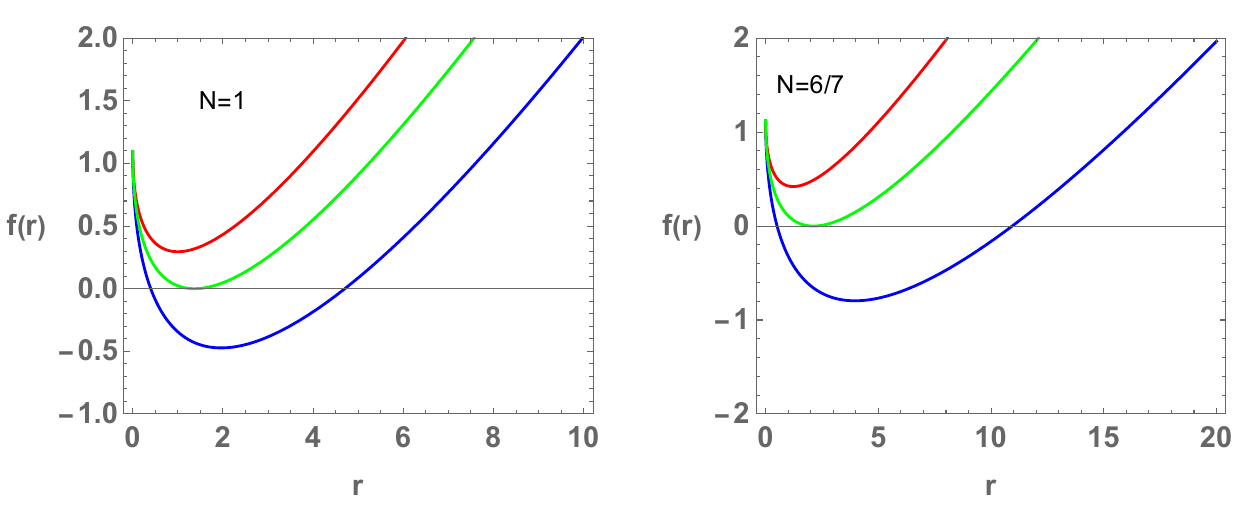}  

\caption{The figure shows  $f(r)$ vs $r$   for $ N = 1$ and $N = \frac{6}{7}$.  Here $Q = 0.37, \beta = 0.1, \gamma = 1, \Lambda = -1$. $M = 0.14$ (blue),  $ M = 0.10$ (red) and $M = 0.12$ (green) is an extreme black hole.}
\label{frr}
 \end{center}
 \end{figure}

Here we will analyze details of thermodynamics for the black hole with $N = \frac{6}{7}$. Notice that $\Lambda < 0$ for black hole horizons to exist.  As discussed in \cite{Chan:1994qa}, it is possible to choose $M$ and $Q$ such that there are two horizons, degenerate horizons or none. In the extreme case,
\be
Q^2 =  \frac{ 4096}{9261} \left(\frac{M^3}{ \beta^{16/7} \gamma^3 \Lambda^2}  \right) = a M^3  \, .
\en
Here,
\be
a = \frac{ 4096}{9261} \left(\frac{1}{\beta^{16/7} \gamma^3 \Lambda^2} \right)  \, .
\en
If $Q^2 < a M^3$,  one will have a black hole solution and if $Q^2 > a M^3$, then one will have a naked singularity. In Fig.$\refb{frr}$ we have plotted $f(r)$ for three  values of $M$ for fixed $Q$. Notice that the relation given for extreme black hole  is different from the one given by Chan and Mann \cite{Chan:1994qa} since we have redefined mass $M$ in this paper.
\subsection{Thermodynamical quantities}

By replacing $N=6/7$ in the quantities obtained in subsection~\ref{First-law-N}, one can obtain the temperature $T$ , volume $V$ and  the horizon radius for the extreme black hole $r_{ex}$ which is tabulated in Table 1  as follows:

\begin{center}
\begin{tabu}{|l|l|l|l|r} \hline \hline
\rowfont{\color{black}}
N  &  Temperature  &    ($r_{ex}$) & Volume     \\ \hline
\rowfont{\color{black}}
$2/3 < N < 2$ & $\frac{-\Lambda \beta ^{2-N}  r_+^N - Q^2}{N \pi  r_+} $ & 
$\left(\frac{ Q^2}{  - \Lambda \beta^{2 - N}} \right)^{1/N} $ &   $\frac{4 \pi  \gamma \, \beta^{2 - N} r_+^{\frac{3 N}{2}-1}}{3 N - 2}$ \\ \hline
\rowfont{\color{black}}
6/7 & $  \frac{-7( Q^2 + r^{6/7} \beta^{8/7} \Lambda)}{ 6 \pi r_+}$ & $\left( \frac{-Q^2}{ \beta^{8/7} \Lambda} \right)^{7/6}$ & $7 \pi \beta^{8/7} \gamma r_+^{2/7}$ \\ \hline
\rowfont{\color{black}}
1  &  $ - \frac{Q^2}{\pi \, r_+}  - \frac{\Lambda \beta}{\pi} $ & 
$\left(\frac{ Q^2}{  - \Lambda \beta} \right) $ & $4 \pi \beta \gamma \sqrt{r_+}$\\ \hline

\rowfont{\color{black}}
2/3 &  $-\frac{3 }{2 \pi r_+} \left( \beta ^{4/3} \Lambda  r_+^{2/3} +  Q^2 \right)$ & $\left( \frac{ Q^2}{ - \beta^{4/3} \Lambda} \right)^{3/2}$ & $2 \pi  \gamma \beta^{4/3} \log \left(\frac{r_+}{\beta }\right)$\\ \hline \hline
\end{tabu}

\vspace{0.3 cm}
{Table 1: Expressions for $T$, $r_{ex}$  and $V$ for all $N$ values}
\end{center}

\begin{figure} [H]
\begin{center}
\includegraphics[scale=0.80]{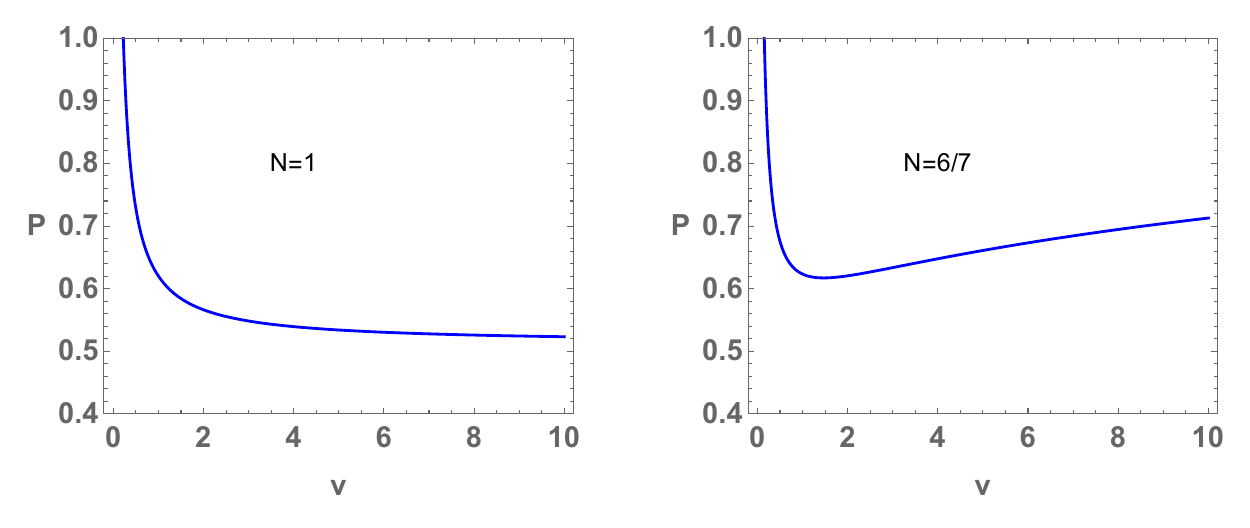}  

\caption{The figure shows  the pressure $P$  vs the specific volume $v$ for $N = 1$ and $N=6/7$. Here $Q = 0.26, T = 0.41$ and $\beta = 0.1$.}
\label{pvsrn}
 \end{center}
 \end{figure}

From eq.$\refb{state}$, the pressure $P$ for $N=6/7$ is given by,

\be \label{pvequation}
P = \frac{1}{\beta^{8/7} 2^{2/7}} \left( \frac{3 T v^{1/7}}{ 28} + \frac{ Q^2}{ 2 \pi v^{6/7}} \right)
\en
In order to study criticality of this thermodynamical system, one needs to find critical points. Such points occur at the inflection points of the $P - v$ diagram (or $P - r_+)$ diagram. The condition for the inflection point is,
\begin{equation}
\left(\frac{\partial P}{ \partial v}\right)_T =  \left(\frac{\partial^2 P}{ \partial v^2}\right)_T = 0
\end{equation}
From eq.$\refb{pvequation}$, $\frac{\partial P}{ \partial v} = 0$ is at $ v = 28Q^2/ ( \pi T)$ and $\frac{\partial^2 P}{ \partial v^2} = 0$ is at $ v = 182^2/ ( \pi T)$.  There are no solutions for $v$ when $\frac{\partial P}{ \partial v} = \frac{\partial^2 P}{ \partial v^2}=0$. Hence, there are no critical points for this case.  $P$ has a minima at $v =  28Q^2/ ( \pi T)$ or at $r_+ = 7 Q^2/ ( \pi T)$.
The pressure  $P$ vs $v$ is plotted in Fig.$\refb{pvsrn}$. As is clear from the figure, there is a minima for pressure at $v = \frac{28 Q^2}{ \pi T}$. Hence for a given black hole,
\be
P >  \frac{1}{8 \beta^{8/7}} \left(\frac{ 7}{ \pi}\right)^{1/7} \left( \frac{Q^2}{T}\right)^{1/7}  \, .
\en


\subsection{ Thermodynamic stability}
\begin{figure} [H]
\begin{center}
\includegraphics[scale=0.80]{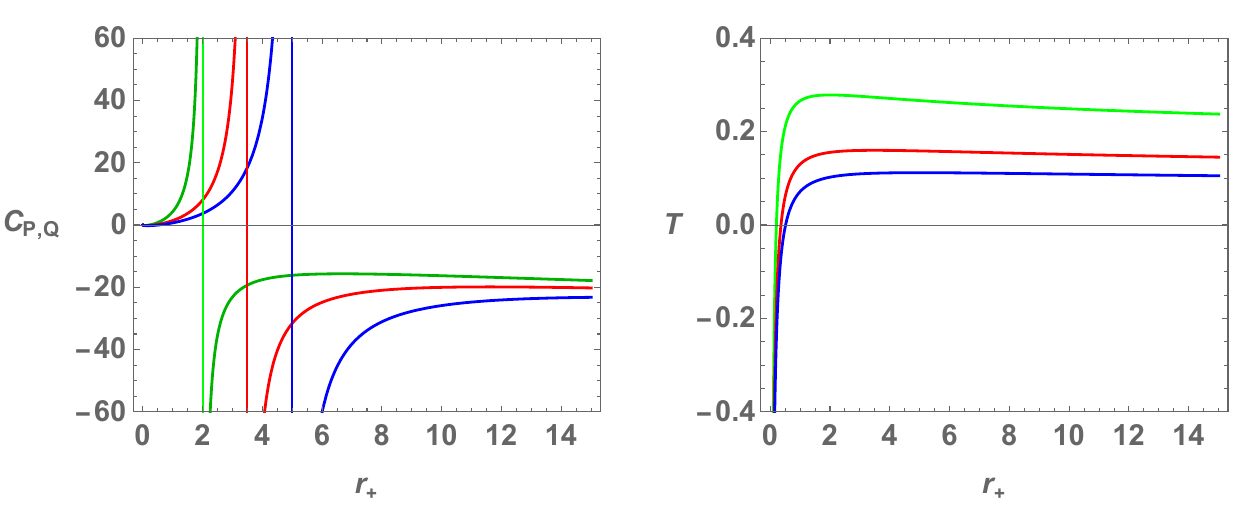}  
\caption{The figure shows   $C_{P,Q}$ and  $T$ vs  $r_+$. Here $N = 6/7, Q = 0.5$, $\Lambda = -1, \gamma =1$, and $\beta =$ 0.97, 0.64, 0.49 (green, red and blue).}
\label{cptn1}
 \end{center}
 \end{figure}

\begin{figure} [H]
\begin{center}
\includegraphics[scale=0.80]{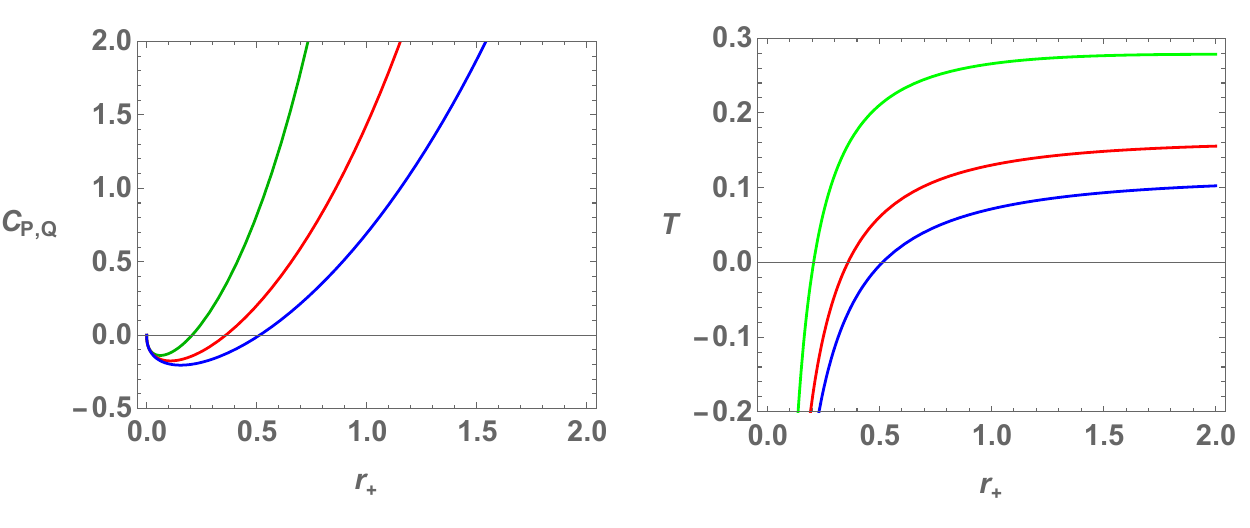}  

\caption{The figure shows    $C_{P,Q}$ and  $T$ vs  $r_+$ with amplified scale close to $r_+ = 0$ for the same cases as in Fig.$\refb{cptn1}.$}
\label{cptn2}
 \end{center}
 \end{figure}

For the black hole to be locally stable, $C_{P,Q} >0$.  From eq.$\refb{C-N}$, one can substitute $N = 6/7$ to obtain the corresponding $C_{P,Q}$ as follow:
\begin{equation}
C_{{P,Q}} =  
\frac{3 \pi  \gamma \,   r^{3/7}_+ \left( \Lambda \beta^{8/7}  r^{6/7}_+  + Q^2\right)}{  2(\Lambda \beta^{8/7} r^{6/7}_+ + 7 Q^2)} 
\label{C67} 
\end{equation} 
We have plotted $C_{P,Q}$ vs $r_+$ and $T$ vs $r_+$ in Fig.$\refb{cptn1}$ and Fig.$\refb{cptn2}$.  Both $T$ and $C_{P,Q}=0$  when the black hole is extreme at $r_+ = r_{ex}$ given in Table 1.

Temperature $T$ has a local maxima at $r_{max}$ given by,
\be \label{tmax}
r_{max} = 7^{(7/6)} \left( \frac{-Q^2}{ \beta^{8/7} \Lambda} \right)^{7/6} 
\en
$C_{P,Q} \rightarrow \infty $ when $r = r_{max}$ and beyond that $C_{P,Q} <0$. Hence, $C_{P,Q} > 0$ between $r_{ex}$ and $r_{max}$ and black holes are locally stable in this region. Hence, small black holes are locally stable and large black holes are locally unstable.

\subsubsection{Absence of  van der Waals, Hawking-page  and Hawking-Page reentrant phase transitions} Our goal here is to understand if there are phase transitions within black hole phases (such as between large and small or between large and thermal-AdS space). When there are multiple phases (such as large, small or thermal-AdS space), the preferred state would be the one with the lowest free energy. In a well known paper, Hawking and Page studied the 4D Schwarzschild-AdS black hole and demonstrated that the large black holes are thermodynamically stable and that there is a phase transition between large black hole and thermal-AdS space (known as Hawking-Page phase transition (HP)) \cite{hawking2}.

\begin{figure} [H]
\begin{center}
\includegraphics[scale=0.80]{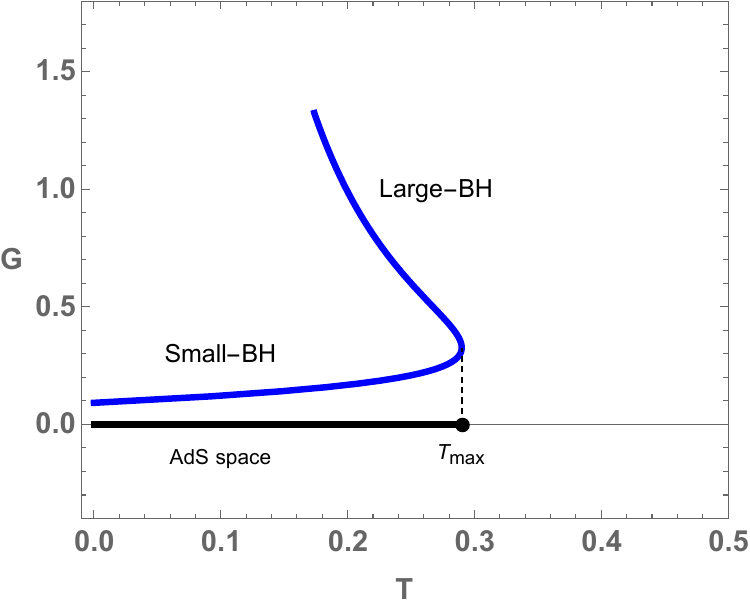}  

\caption{The figure shows  the  Gibbs free energy $G$  vs the temperature $T$ for fixed charge. Here $N = \frac{6}{7}, Q = 0.5$, $\Lambda = -1, \gamma=1$,  and $\beta = 1$.
}
\label{gvstn}
 \end{center}
 \end{figure}

\begin{figure} [H]
\begin{center}
\includegraphics[scale=0.70]{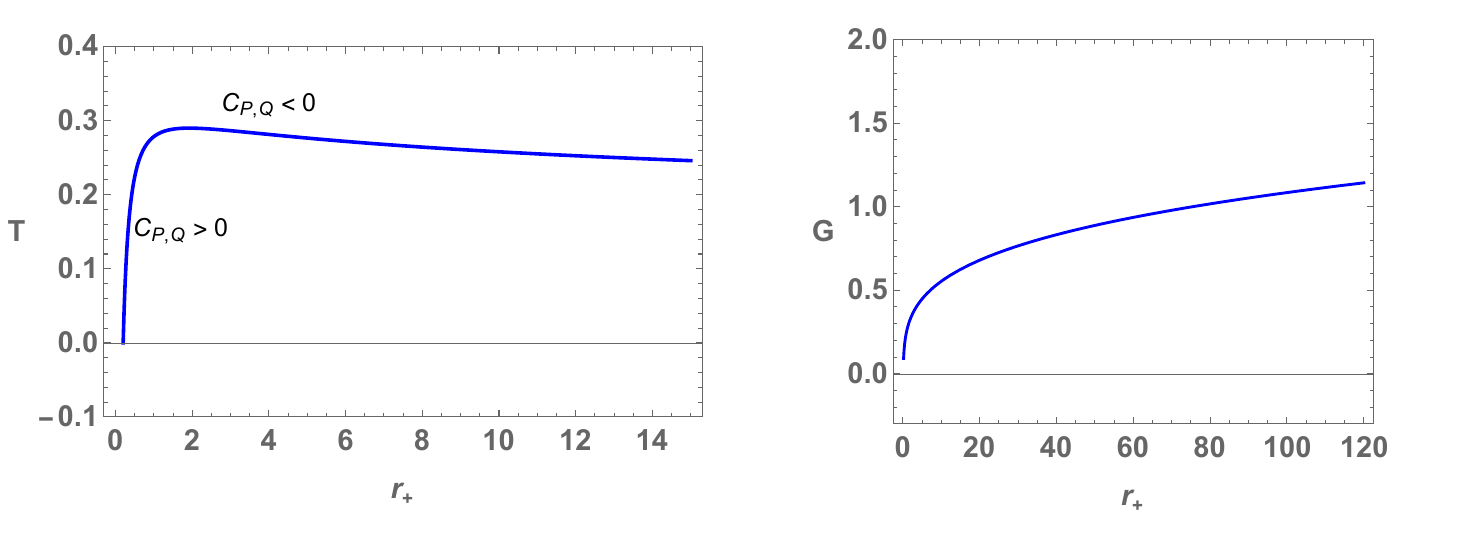}  

\caption{The figure shows   the temperature $T$ vs $r_+$ and Gibbs free energy $G$  vs $r_+$ for fixed charge. Here $N = \frac{6}{7}, Q = 0.5$, $\Lambda = -1, \gamma=1$, and $\beta = 1$.
}
\label{gandtn}
 \end{center}
 \end{figure}

To understand global stability of a black hole one has to study Gibbs free energy. We have substituted $N = 6/7$ to eq.$\refb{gibbs-N}$ to obtain $G$ for $N = 6/7$ and plotted $G$ vs $T$ in  Fig.$\refb{gvstn}$. We have plotted $G$ vs $r_+$ and $T$ vs $r_+$ in Fig.$\refb{gandtn}$.
As demonstrated in Fig.$\refb{tvsn}$ and eq.$\refb{tmax}$, the temperature $T$ has a maximum, $T_{max}$ at $r = r_{max}$. When $T > T_{max}$, no black holes can exist. For $T < T_{max}$ there are two branches of black holes: one corresponds to small black hole with smaller (positive) Gibbs free energy and the other corresponds to large black hole with large, still positive, free energy. Smaller black holes are locally stable and larger black holes are locally unstable. It is clear that the thermal-AdS space has lower $G$ than the small black holes. However, since we are studying the charged black holes in the canonical ensemble where the charge $Q$ is fixed, the thermal-AdS space is not accessible to the charged black hole: the black hole cannot evaporate completely while keeping the charge Q non-zero \cite{marks}\cite{anabalon}. Hence the lowest free energy state is the small black hole and is the preferred state. Hence it is not possible for a charged dilaton black hole to have HP transition in the canonical ensemble. We will discuss HP transitions for uncharged dilaton black holes in  Appendix A and in the grand canonical ensemble in section 8.

This behavior of the Gibbs free energy of $N=6/7$ black hole is  similar to the Reissner-Nordstr\"om black hole in 4 dimensions, as shown in \cite{mann1}. There,  the smaller, strongly charged black hole  is thermodynamically preferred over the larger, weakly charged black hole. The Gibbs free energy of the Kerr black hole, considered in canonical ensemble with the angular momentum $J$ replaced with charge $Q$, also behaved  similar to the RN black hole and the dilaton black hole discussed above \cite{mann1}. 

The charged dilaton black hole does not have HP phase transitions (as discussed above) or van der Waals type phase transition (since there are no critical points) as found in some charged AdS black holes \cite{chamblin}\cite{daripa}. 

Re-entrant Hawking-Page phase transitions occurs when there are two HP phase transitions with two different temperatures. When the temperature is increased gradually, there could be small black hole/thermal-AdS HP transition and then thermal-AdS/large black hole HP transition which may happen successively \cite{cui}\cite{marks}. Since for charged dilaton black hole there are no HP transitions, one could conclude that there are no re-entrant HP transitions for the above charged black holes.

\section{Thermodynamics of the black holes  for $N=1$ case}
We will study $N=1$ case in order to quantitatively understand the thermodynamics of black holes in the region, $1 \leq N < 2$. $N=1$ black hole is also important given that it is related to string theory. 


\subsection{Horizon structure}
The metric of the black hole for $N=1$ is given in eq.$\refb{metricn1c}$.  The black hole has two horizons when  the mass $M > \gamma Q \sqrt{- \beta \Lambda}$ and degenerate horizon {{$r_{ex}$} given in Table: 1  when $M = \gamma Q \sqrt{- \beta \Lambda}$

\subsection{Thermodynamical quantities }

By replacing $N=1$ in the quantities obtained in subsection~\ref{First-law-N}, one can obtain the emperature $T$ , volume $V$ and  $r_{ex}$ which is tabulated in Table 1.

Since the duality between the N=1 charged dilaton black hole and the uncharged dilaton black hole was explained in the introduction, some comments about the behavior of thermodynamical quantities under duality transformations are in order: from the expressions in Table 1, it is clear that when $Q =0$, they differ with the ones with $Q \neq 0$. Hence it is clear that $T, S, V$ are not invariant under duality.

Notice that when the black hole is extreme, then $T =0$. Also when $r_+ \ra \infty$, $T \ra -\frac{ \Lambda \beta}{ \pi}$. There are no local minima or maxima for $T$. Hence black holes cannot exist for $T <0$ or $T > -\frac{ \Lambda \beta}{ \pi}$. Fig.$\refb{cpt}$ shows the behavior of $T$ vs $r_+$ for $N=1$ case.

We also want to note that the volume $V$ given in Table 1 is different from the geometric volume $\pi r_+^2$.  The charge $Q$ is related to the electric charge $Q_{BTZ}$ of the BTZ black hole \cite{Banados:1992wn, Martinez:1999qi} as follows from Ref.~\cite{Dehghani:2019xhm}
\begin{equation} 
Q_{BTZ} =  \frac{\gamma \, Q}{2}   \, .
\label{}
\end{equation}
In order to study critical points, one can consider the pressure P given by,
\begin{equation}
P = \frac{ T}{ 8 \beta} + \frac{Q^2}{ 2 \pi \beta v} \,   
\end{equation}
From above, $\frac{\partial P}{ \partial v} = -Q^2/( 2 \pi v^2 \beta)$ and $\frac{\partial^2 P}{ \partial v^2} = Q^2/ ( \pi v^3 \beta)$. There are no solutions for $v$ when $\frac{\partial P}{ \partial v} = \frac{\partial^2 P}{ \partial v^2}=0$. Hence, there are no critical points for this case.  $P$ is a monotonically decreasing function of $v$. The pressure $P$  vs $v$ is plotted in Fig.$\refb{pvsrn}$  to demonstrate that there are no critical points.


\subsubsection{Thermodynamic stability}
From eq.$\refb{C-N}$, the specific heat capacity $C_{P,Q}$ is given by,
\be \label{cp}
C_{P,Q}  = \frac{ \pi \gamma \sqrt{r_+} ( -Q^2 + 8 \pi \beta P r_+ ) }{ 4 Q^2 \beta}
\en
For the black hole to be locally stable, $C_{P,Q} >0$. From eq.$\refb{cp}$ this implies,
\be
P > \frac{Q^2}{ 8 \pi \beta r_+} \Rightarrow
- \Lambda > \frac{Q^2}{ \beta r_+} 
\en
Hence it is clear that if the black hole to be locally stable, then $\Lambda$ has to be negative or has or be in anti-de Sitter space. This is very similar to the Schwarzschild anti-de Sitter black hole where it was demonstrated that $-\Lambda r_+^2 > 1$ for $C_p$ to be positive \cite{dolan}. If we rewrite $C_{P,Q}$ in terms of $T$, it is given by,
\be
C_{P,Q} = \left( \frac{ \pi^2 r_+^{3/2}}{ 4 Q^2 \beta} \right) T
\en
Since $\beta >0$, it is clear that when $ T >0$, $C_{P,Q} >0$. $C_{P,Q} = T=0$  when the black hole is extreme with $r_{ex} = -\frac{Q^2}{ \beta \Lambda}$. When $r \ra \infty $, $T \ra = -\frac{\Lambda \beta} {\pi}$. Hence for large $r_+$, $C_{P,Q} \ra \infty$. Hence for all physical black holes with temperature between $0$ and $-\frac{Q^2}{\beta \Lambda} $, $C_{P,Q} \geq 0$. When $C_{P,Q}$ and $T$ are plotted against $r_+$, it is clear that both are positive for the same values of $r_+$ as demonstrated in Fig.$\refb{cpt}$. Hence, provided $\Lambda < 0$, all black holes with $N=1$ are locally stable.

\begin{figure} [H]
\begin{center}
\includegraphics[scale=0.80]{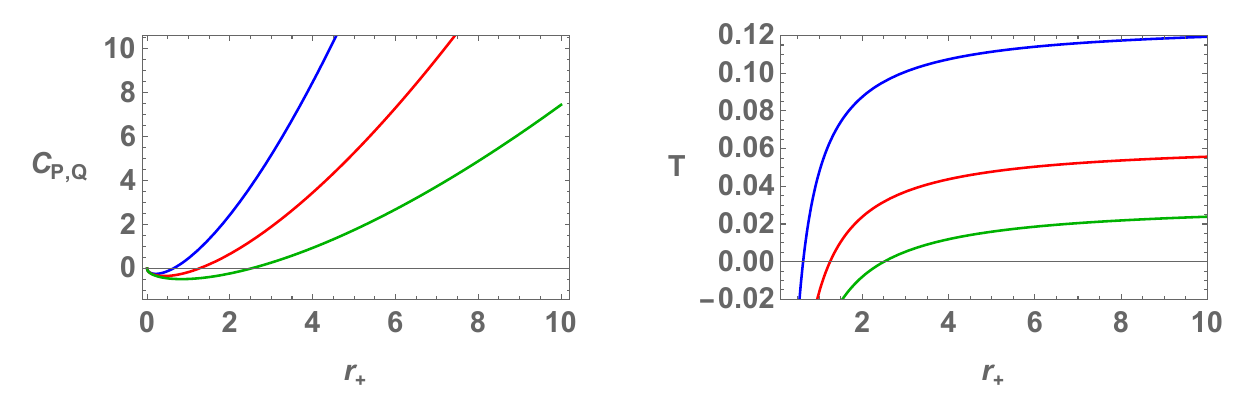  }
\caption{The figure shows  $C_{P,Q}$ and  $T$ vs  $r_+$ for   $N = 1$. Here $Q = 0.5,$ $\Lambda = -0.2, \gamma =1$ and $\beta = 0.5,1,2$ (green, red, blue).}
\label{cpt}
 \end{center}
 \end{figure}
In order to study global stability of the black hole, we will once again substitute $N=1$ to eq.$\refb{gibbs-N}$ to find $G$. Note that the Gibbs free energy does not depend on $\Lambda$ as follows:
\begin{equation}
G = \frac{\gamma Q^2}{\sqrt{r_+}}
\end{equation} 
and,
\be
\frac{d G}{ dr_+} = - \frac{ \gamma  Q^2}{ 2 (r_+)^{3/2}} < 0 \, .
\en
In terns of $T$, Gibbs free energy can be written as,
\be
G = \gamma Q \sqrt{ - ( T \pi + \Lambda \beta)}
\en
When $T=0$, $G = \gamma Q \sqrt{( - \Lambda \beta)} >0$. Since $\frac{d G}{ dr_+} <0$, $G$ is  decreasing and it does not have a local minimum. Also, $G >0$ for all black holes and will approach zero  when $ r_+ \ra \infty$ or $ T \rightarrow  -\frac{ \Lambda \beta} {\pi}$. $G$ has only one branch. Hence we can conclude that the black hole does not admit  van der Waals type behavior for $N=1$ black holes. The graph of G vs T is given in Fig.$\ref{GvsT1}$. From the graph it is clear that the thermal-AdS space has lower free energy, however, since the black hole is charged, the black hole  is the preferred state. There are no HP  phase transitions.

\begin{figure} [H]
\begin{center}
\includegraphics[scale=0.80]{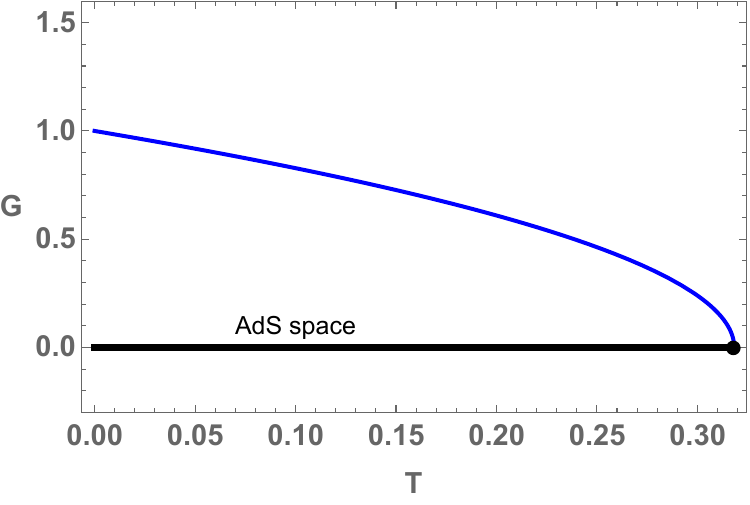}
\caption{The figure shows  G vs T for   $N = 1$ black hole. Here $Q = 1,$ $\Lambda = -1, \gamma =1$ and $\beta = 1$.}
\label{GvsT1}
 \end{center}
 \end{figure}


\section{Thermodynamics for the $N = \frac{2}{3}$ case}

\subsection{Horizon structure}

When we set $N = 2/3$ in the metric function given by eq.$\refb{metric-f-N}$, we find that it exhibits a singularity. Numerous studies~\cite{Dehghani:2017zkm, Xu:2019pap, Dehghani:2018lat, Dehghani:2018hpb, Dehghani:2019dab, Dehghani:2019noj, Dehghani:2022ntd, Dehghani:2023iau, Dehghani:2023zga} have explored models of charged black holes in three-dimensional Einstein-Dilaton gravity. In these studies, cases leading to singularities are individually incorporated into the general expression for the metric function. For the dilaton black holes considered in this paper, a detailed study for $N= 2/3$ is done by Xu in reference ~\cite{Xu:2019pap}. We will adopt the metric function for $N = 2/3$ derived by Xu ~\cite{Xu:2019pap} as follows:
\begin{equation} 
f(r) =  -\frac{24 M r^{2/3}}{\gamma} - 6 \beta^{4/3} \Lambda \, r^{2/3} \log \left(\frac{r}{\beta }\right)  + 9 Q^2  \, .
\label{}
\end{equation}
Fig.$\refb{f-23}$ illustrates the behavior of this metric function with respect to the radial coordinate.

\begin{figure} [H]
\begin{center}
\includegraphics[scale=0.80]{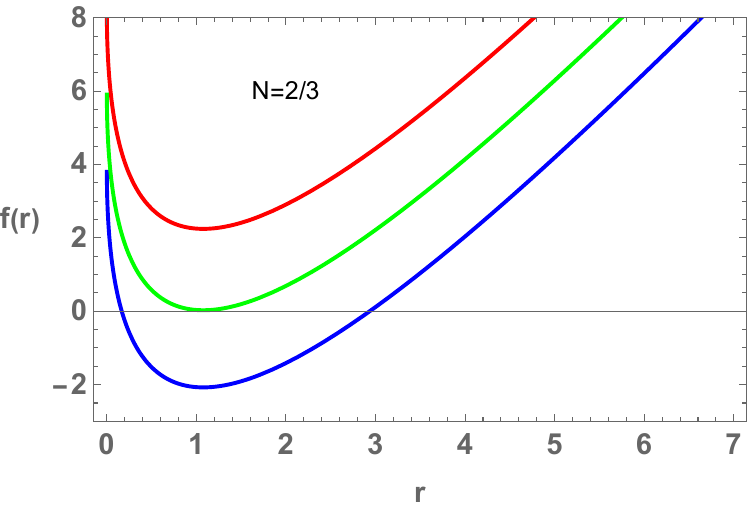}  

\caption{The figure shows $f(r)$ vs $r$ for $N=2/3$. Here  $M = 0.3$, $\gamma = 1$,  $\beta = 0.7$, and $\Lambda = -1$.  $Q = 0.95$ (red), $Q=0.65$ (blue), and $Q=0.81$ (green) is an extreme black hole.}

\label{f-23}
 \end{center}
 \end{figure}
 By considering $f(r_+) = 0$, we can obtain
\begin{equation}
M = \frac{3 \gamma Q^2}{8 r_+^{2/3}}-\frac{1}{4} \gamma \beta^{4/3} \Lambda  \log \left(\frac{r_+}{\beta }\right) \, .
\label{}
\end{equation}
\subsection{First Law of thermodynamics for $N=\frac{2}{3}$ case}
In this case the entropy is given by, 
\begin{equation} \label{entropy23}
S = \frac{2 \pi r_+ R(r_+)}{4} = \frac{\pi}{2} \gamma \, r_+^{1/3}    \, .
\label{}
\end{equation}
The quantities $S$, $P$, $Q$ and $\beta$ are related to the mass $M$ as follows
\begin{equation}
M = \pi  \gamma \left[2 \beta^{4/3} P \log \left(\frac{8 S^3}{\pi ^3 \beta  \gamma ^3}\right) + \frac{3 \pi \gamma^2 Q^2}{32 S^2}\right]   \, .
\label{}
\end{equation}
Similar to  $\frac{2}{3} < N<2$ cases, the first law of black hole thermodynamics for $N = \frac{2}{3}$  takes the form
\begin{equation}
dM = T dS + V dP + \Phi dQ + \Omega d\beta 
\,\,\label{} \,  ,
\end{equation}
where, in the present case, the temperature is given by the expression
\begin{equation} \label{temp23}
T = \left(\frac{\partial M}{\partial S}\right)_{P,Q,\beta}  = -\frac{3 }{2 \pi r_+} \left( \beta ^{4/3} \Lambda  r_+^{2/3} +  Q^2 \right) \, .
\label{}
\end{equation}
We get the same result if we use the surface gravity 
\begin{eqnarray} \label{temp5}
 T_{H}   &=&  \frac{1}{4 \pi} \left|\frac{dg_{tt}}{dt} \sqrt{-g^{tt} g^{rr}}  \right|_{r_+}  = -\frac{1}{2 \pi  (\beta^2 r_+)^{1/3}}\left(8 M + 3 \beta ^2 \Lambda  + 2 \beta ^2 \Lambda  \log \left(\frac{r_+}{\beta }\right)\right)  \nonumber  \\
 &=&  -\frac{3 }{2 \pi r_+} \left( \beta ^{4/3} \Lambda  r_+^{2/3} +  Q^2 \right) 
  \, .
\label{}
\end{eqnarray}
If $ P = -\frac{\Lambda}{ 8 \pi}$, then one can obtain the thermodynamic volume as,
\begin{equation} \label{volume23}
V = \left(\frac{\partial M}{\partial P}\right)_{S,Q,\beta}  = 2 \pi  \gamma \beta^{4/3} \log \left(\frac{r_+}{\beta }\right) \, .
\label{}
\end{equation}
Notice that the thermodynamics volume differs from the geometric volume as for other $N$ values. The expression for the electric potential at the horizon gives
\begin{equation} 
\Phi = \left(\frac{\partial M}{\partial Q}\right)_{S,P,\beta}  = \frac{3 \gamma Q}{4 r_+^{2/3}}  \, .
\label{}
\end{equation}
The thermodynamic quantity $\Omega$ conjugate to the parameter $\beta$ is given by,
\begin{equation} 
\Omega = \left(\frac{\partial M}{\partial \beta}\right)_{S,P,Q}  = 
\frac{1}{12}\gamma  \beta^{1/3}  \Lambda  \left[3 - 4 \log \left(\frac{r_+}{\beta} \right)\right]
\, .
\label{}
\end{equation}

We have summarized the temperature $T$ , volume $V$ and  the horizon radius for the extreme black hole $r_{ex}$  in Table 1.
In order to study critical points, one can substitute $\Lambda = - 8 \pi P$ in eq.$\refb{temp5}$, and $v = 4 r_+$, and write  the state equation as,

\be
P = \frac{Q^2}{ 2^{5/3} \pi \beta^{4/3}v^{2/3}} + \frac{ v^{1/3} T}{ 2^{5/3} \beta^{4/3}}
\en
From above, $\frac{\partial P}{ \partial v} = 0$ is at $ v = 4Q^2/ ( \pi T)$ and $\frac{\partial^2 P}{ \partial v^2} = 0$ is at $ v = 10Q^2/ ( \pi T)$.  There are no solutions for $v$ when $\frac{\partial P}{ \partial v} = \frac{\partial^2 P}{ \partial v^2}=0$. Hence, there are no critical points for this case.  $P$ has a minima at $v =  4Q^2/ ( \pi T)$ or at $r_+ = Q^2/ ( \pi T)$.
The graph for $P$ vs $v$ is given in Fig.$\refb{pvst23}$. One can observe that there is a minimum   similar to the $N = 6/7$ case.
\begin{figure} [H]
\begin{center}
\includegraphics[scale=0.80]{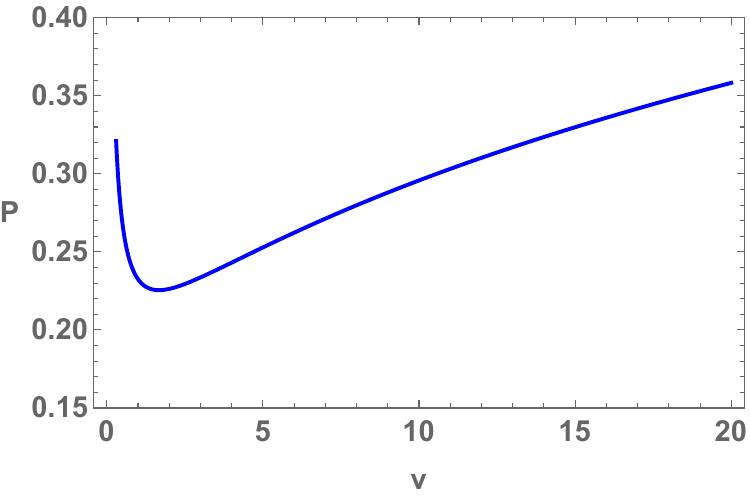}  

\caption{ P vs $r_+$ for $N=2/3$ case. Here $Q =1; \beta =1$ and $T = 0.5$}
\label{pvst23}
 \end{center}
 \end{figure}
\subsection{Thermodynamics stability for $N= 2/3$ case}

In this section we will analyze details of thermodynamics for the black hole with $N=2/3$. As seen in Fig.$\refb{f-23}$, this black hole can have two horizons or degenerate horizons. When the black hole has degenerate horizons, $f(r)= f'(r) =0$ leading to the temperature also being zero. From eq.$\refb{temp5}$, $T=0$ when,
\be \label{rex23}
r_{ex} = \left( \frac{ Q^2}{ - \beta^{4/3} \Lambda} \right)^{3/2}
\en
Also, the temperature has a maximum when the horizon radius is,
\be
r_{max} = \frac{ 3 \sqrt{3} Q^3}{ \beta^2 (-\Lambda)^{3/2}}
\en
Temperature vs $r_+$ and $C_{P,Q}$ vs $r_+$  is plotted in Fig$\refb{cvt4}$. One can notice that there is indeed a maximum for the temperature. Hence, black holes can exist only below the maximum temperature given by,
\be
T_{max} =  \frac{\beta^2 (- \Lambda)^{3/2} }{ \sqrt{3} \pi Q}
\en
The specific heat capacity, $C_{P,Q}$ is given by,

\begin{equation}
C_{P,Q} = -\frac{\pi  \gamma r_+^{1/3}\left(Q^2  + \beta^{4/3} \Lambda  r_+^{2/3}\right)}{2( 3 Q^2 +  \beta^{4/3} \Lambda r_+^{2/3})}
\,\,\label{} \,  .
\end{equation}
For a black holes to be stable locally, $C_{P,Q} >0$. From the Fig.$\refb{cvt4}$ and Fig.$\refb{cvt5}$, both T and $C_{P,Q}$ are zero when the black hole is extreme at $r_+ = r_{ex}$. $C_{P,Q} \rightarrow \infty$ when $r_+ = r_{max}$ and beyond that $C_{P,Q} <0$. Hence $C_{P,Q} >0$ only between $r_{ex}$ and $r_{max}$ and that is the only region black hols are locally stable.

\begin{figure} [H]
\begin{center}
\includegraphics[scale=0.80]{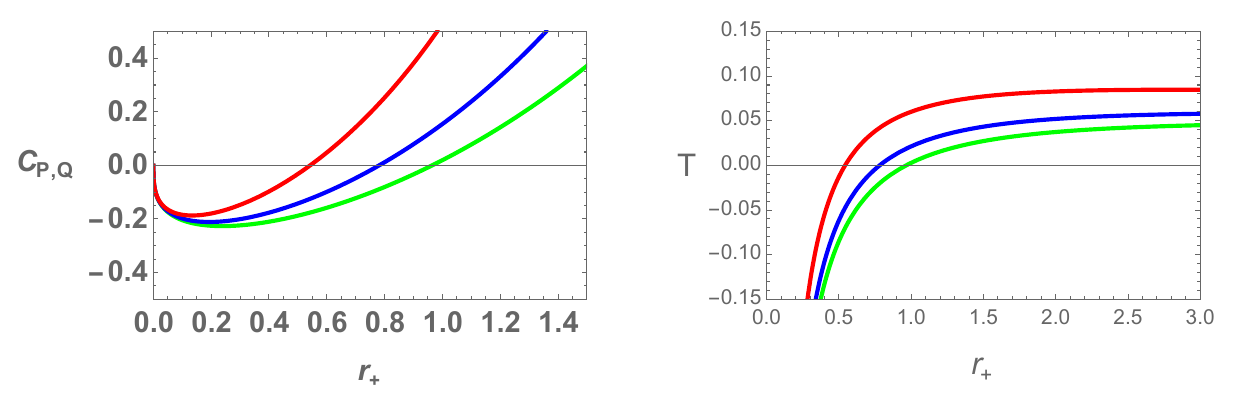}  

\caption{$C_{P,Q}$ and  $T$ vs  $r_+$ for $N=2/3$ black hole for fixed charge. Here, $Q = 0.5, \Lambda = -1, \gamma = 1$ and $\beta = 0.48, 0.40, 0.36$ (red, blue, green).}
\label{cvt4}
 \end{center}
 \end{figure}

\begin{figure} [H]
\begin{center}
\includegraphics[scale=0.80]{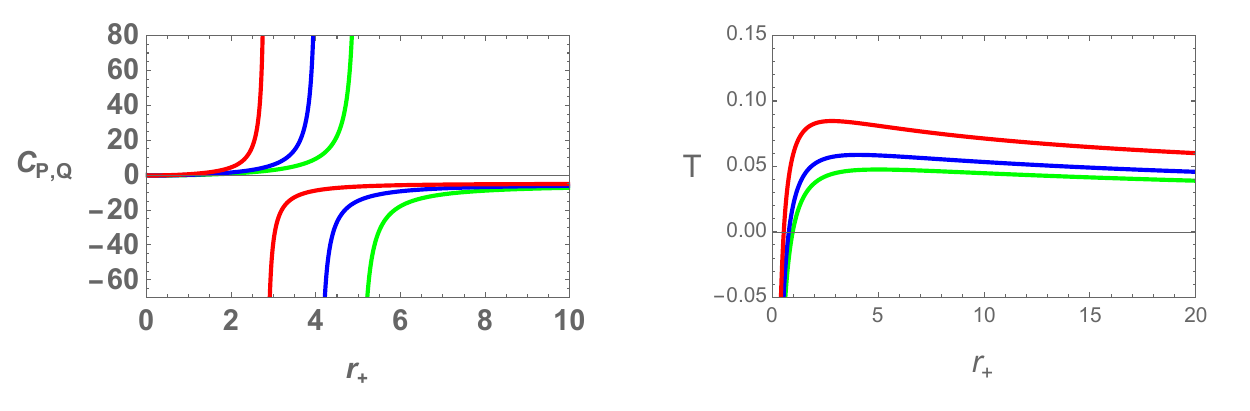}  

\caption{ $C_{P,Q}$ and  $T$ vs $r_+$  with amplified scale close to $r_+=0$ for the same cases as in Fig.$\refb{cvt4}$}
\label{cvt5}
 \end{center}
 \end{figure}
 To understand the global stability one has to study the Gibbs free  energy as we did for other cases of $N$. The Gibbs free energy $G = M - TS$ is given as,
\begin{equation}
G = \frac{3 \gamma \beta^{4/3} \Lambda }{4} -\frac{1}{4} \gamma \beta^{4/3} \Lambda  \log \left(\frac{r_+}{\beta }\right) + \frac{9 \gamma Q^2}{8 r_+^{2/3}} 
\,\,\label{} \,  .
\end{equation}
In Fig.$\refb{gvst3}$ we have plotted G vs T and in Fig.$\refb{gandt3}$ G vs $r_+$ and T vs $r_+$. Similar to the $N = 6/7$ case, for $T < T_{max}$ there are two branches of black holes: small and large. Smaller black holes are locally stable and preferred globally since it has lower values of $G$. There are no black holes beyond $T = T_{max}$. There are no first-order phase transitions such as the 4D Schwarzschild-AdS black hole because it is charged \cite{hawking2} or van der Waals phase transitions such as for the 4D Reissner-Nordstr\"om-AdS black hole \cite{mann3}.

\begin{figure} [H]
\begin{center}
\includegraphics[scale=0.80]{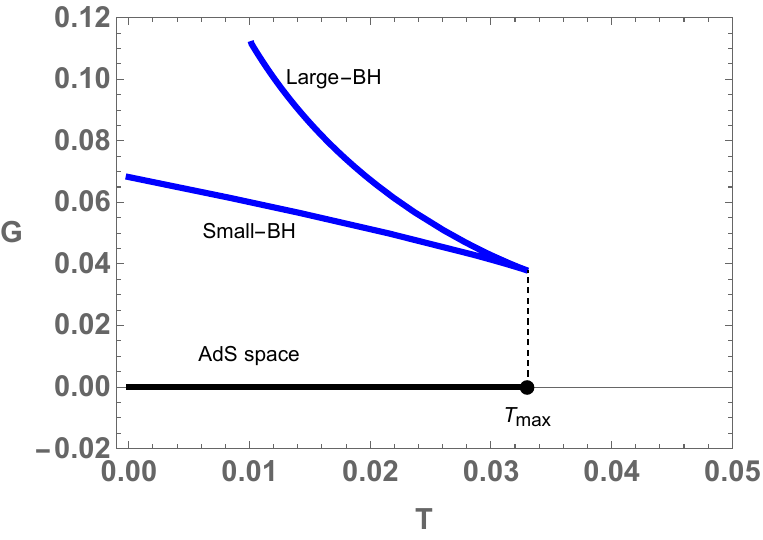}  

\caption{Gibbs free energy $G$ is represented as a function of horizon $r_+$, for fixed charge for $N = \frac{2}{3}$. Here, $Q =0.5, \beta = 0.3, \gamma = 0.45, \Lambda = -1$.}
\label{gvst3}
 \end{center}
 \end{figure}

\begin{figure} [H]
\begin{center}
\includegraphics[scale=0.80]{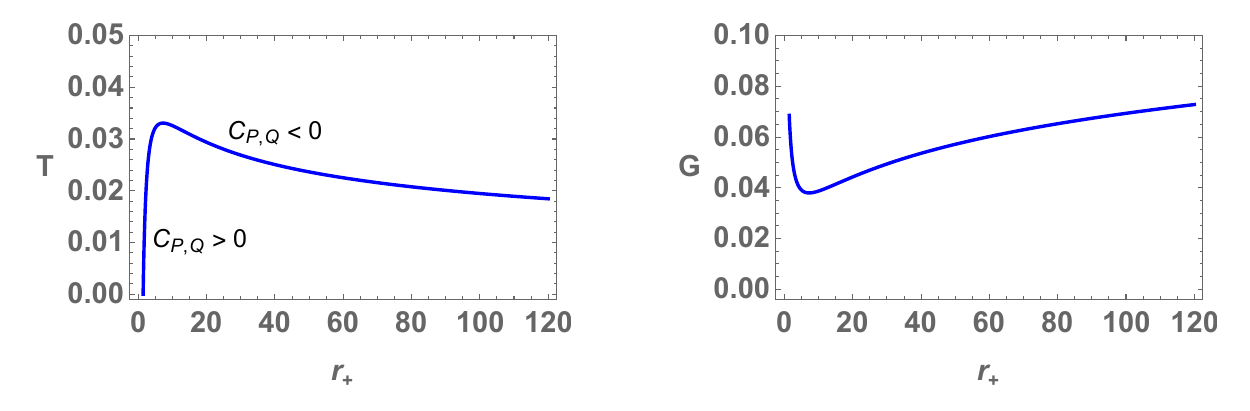}  
\caption{ The figure shows   the temperature $T$ vs $r_+$ and Gibbs free energy $G$  vs $r_+$ for fixed charge for $N=\frac{2}{3}$. Here $Q =0.5, \beta = 0.3, \gamma = 0.45, \Lambda = -1$.}
\label{gandt3}
 \end{center}
 \end{figure}

\section{Comparison of thermodynamics with the BTZ black holes in the canonical ensemble}

Here we would like to compare  thermodynamic stability of the dilaton black holes with the BTZ (charged and uncharged) black holes. The thermodynamics of the BTZ black hole is addressed in \cite{mann2} and \cite{mann4}. For the sake of completeness, we will summarize some important thermodynamical quantities. First let us  focus on the uncharged static BTZ black hole: the metric of the uncharged static BTZ black holes is:
\be
- (- 8M - r^2 \Lambda)(dt^2 + \frac{dr^2}{(- 8M - r^2 \Lambda)} + r^2 d \theta^2
\en
with
The temperature T, the entropy S, specific heat capacity $C_{P}$ and Gibbs free energy are given by
\be
T = - \frac{ \Lambda r_+}{ 2 \pi}; \hspace{0.4cm} S = \frac{ \pi r_+}{2}; \hspace{0.4 cm} C_P = \frac{ \pi r_+}{2}; \hspace{0.4 cm} G  = \frac{ r_+^2 \Lambda}{8}
\en
Hence for all black holes, both $T$ and $C_P$ are greater than zero. Hence, the black holes are locally stable thermodynamically for all black holes similar to the N=1 dilaton black hole. On the other hand, the Gibbs free energy for the BTZ black hole is smaller than zero and in terms of $T$, it is,
\be
G = \frac{ \pi^2 T^2}{2 \Lambda} 
\en
Hence, $G$ is proportional to $T^2$ and when $G$ is plotted against $T$, there won't be swallow tail behavior resembling a van der Walls type phase transition. Also, $G < 0$ for all black holes: thermal-AdS space has large free energy, hence all black holes are stable globally and are the preferred state. In fact, this behavior is very similar to the $N=1$ dilaton black hole where all black holes are stable and have one branch for $G$.

Next, we will briefly describe the behavior of the charged BTZ black hole. Thermodynamical quantities are given in \cite{mann4} and we will state them for the sake of completeness:
\be
T = - \frac{ \Lambda r_+}{ 2 \pi} - \frac{Q^2}{ 8 \pi r_+}; \hspace{0.4cm} S = \frac{ \pi r_+}{2}; \hspace{0.4 cm} C_P = -\frac{ \pi r_+}{2} \left( \frac{ Q^2 + 4 r_+^2 \Lambda}{ Q^2 - 4 r_+^2 \Lambda}\right); \hspace{0.4 cm} G = \frac{Q^2}{16} + \frac{r_+^2 \Lambda} { 8} - \frac{Q^2}{16} ln ( r_+ \sqrt{-\Lambda})
\en
It can be observed that $T \geq 0$ only for $r_+ \geq Q/ ( 2 \sqrt{-\Lambda})$ and, $C_P \geq 0$ when $T \geq 0$. Hence,  all physical black holes are thermodynamically stable locally. To understand the global stability, we have mentioned the Gibbs free energy which was  given in \cite{mann4} above. The graph of $G$ vs $T$ (Fig.$\refb{gvstcbtz}$ clearly shows that there are no critical van der Waals behavior as determined in \cite{mann4} by computing $dG/dr_+$. From observing the graph it is clear that $G >0$  until point $L$ (which are small black holes) and only after that it has negative values (for large black holes). Until point $L$, thermal-AdS has lower $G$ than the small black holes. However, since it is charged, as discussed for $N=6/7$ case, thermal-AdS space is not accessible: there are no HP transitions at $L$ and  large black holes are preferred globally.
\begin{figure} [H]
\begin{center}
\includegraphics[scale=0.70]{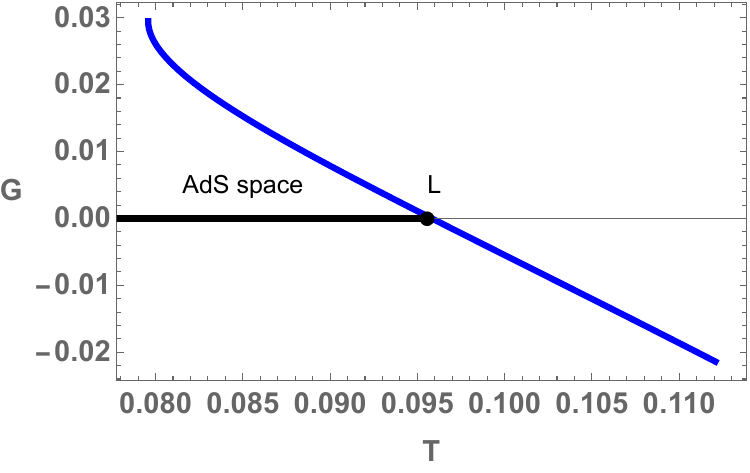}
\caption{The figure shows  G vs T for charged BTZ black hole. Here $Q = 0.5,$ and  $\Lambda = -1$.}
\label{gvstcbtz}
 \end{center}
 \end{figure}{

Hence one can conclude that the thermodynamic behavior of the charged BTZ black hole differs from the N=1, N=2/3 and N=6/7 charged dilaton black hole. The presence of the dilaton has made a difference in thermodynamic stability.


\section{Comments on thermodynamics of the dilaton black holes in the grand canonical ensemble}
To complete our discussion of thermodynamics and phase transitions for the dilaton black holes, we will address thermodynamics in the grand canonical ensemble in this section. There are many papers in the literature on studies of grand canonical ensemble and we will only mention few here \cite{hale}\cite{liang2}\cite{huang2}. In this ensemble,  the electric potential $\Phi$ will be fixed rather than the charge $Q$. The thermodynamic potential $G_{\Phi}$ in grand canonical ensemble is  related to the Gibbs free energy for the canonical ensemble, $G$ via a Legendre transformation given by $G_{\Phi} = G - \Phi Q = M - TS - \Phi Q$. The specific heat would be $C_{\Phi} = T \left(\frac{\partial S}{ \partial T}\right)_{\Phi}$. First, we will study the charged BTZ in grand canonical ensemble as follows.
\subsection{Charged BTZ black hole in grand canonical ensemble}
According to \cite{mann4}, $ \Phi = -\frac{Q}{8} ln( r_+ \sqrt{-\Lambda})$. Hence,
\begin{equation}
    G_{\Phi} = G - \Phi Q =  \frac{Q^2}{16} + \frac{r_+^2 \Lambda} { 8} + \frac{Q^2}{16} ln ( r_+ \sqrt{-\Lambda})
\end{equation}
One can substitute $Q = -8 \Phi/ ln(r_+ \sqrt{-\Lambda})$ to write thermodynamical quantities in terms of the potential $\Phi$, in order to  study the behavior of the black hole for fixed $\Phi$.

\begin{figure} [H]
\begin{center}
\includegraphics[scale=0.80]{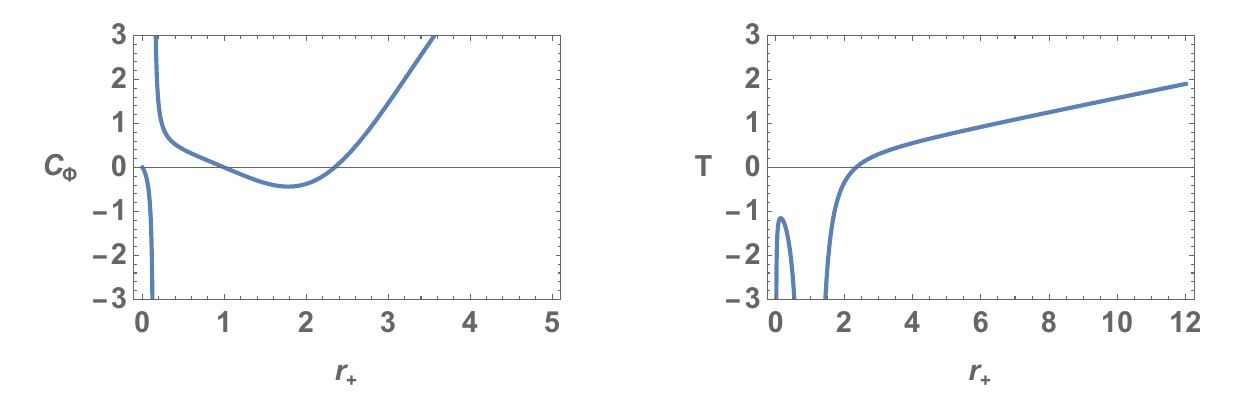}
\caption{The figure shows  $C_{\Phi}$ vs $r_+$ and $T$ vs $r_+$ for charged BTZ black hole. Here $\Phi = 0.5,$ and  $\Lambda = -1$.}
\label{cptgrandBTZ}
 \end{center}
 \end{figure}
From the graphs in Fig.$\refb{cptgrandBTZ}$, one can see that $T >0$ and $C_{\Phi} >0$ for the same $r_+$ value ($r_+$ = 2.35 for the above case). Hence, all physical black holes with positive $T$ are locally stable with positive specific heat capacity. To understand the global stability we have plotted $G_{\Phi}$ vs T in Fig.$\refb{gvstgrandBTZ}$. For $T < T_{HP}$, the radiation phase (characterized by $G_{\Phi}=0$) is preferred thermodynamically since it has lower free energy than  small black holes. For $T > T_{HP}$, the large black holes are preferred since it has lower free energy than the radiation phase. At $T = T_{HP}$, there is an analogue of  Hawking-Page phase transition (HP) between the radiation phase and the large black hole phase.

\begin{figure} [H]
\begin{center}
\includegraphics[scale=0.75]{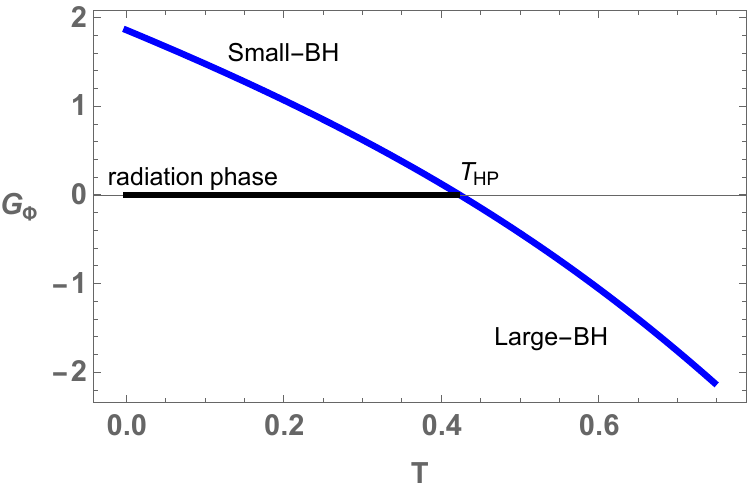}
\caption{The figure shows  $G_{\Phi}$ vs $T$  for charged BTZ black hole in grand canonical ensemble. Here $\Phi = 0.5,$ and  $\Lambda = -1$.}
\label{gvstgrandBTZ}
 \end{center}
 \end{figure}

\subsection{ Charged dilaton black holes in grand canonical ensemble}
For $2/3 < N < 2$, $T, G_{\Phi}, C_{\Phi}$ with the potential $\Phi$ as a constant can be computed by substituting $Q = \frac{( 2 - N) r_+^{1 - N/2} \Phi}{\gamma}$ in eq.$\refb{tempN}$, eq.$\refb{specificN}$ eq.$\refb{gibbsN}$.

For $N=1$, $G_{\Phi}=0$ for all $\Phi$. Also, $T = - \frac{\Phi^2}{\pi \gamma^2} - \frac{\Lambda \beta }{ \pi}$ which is constant for constant $\Phi$. Hence, there are no phase transitions in this case.

For $N=6/7$ case, we observed that the temperature $T$ and  $C_{\Phi}$ behave similar to the canonical case. We have plotted $G_{\Phi}$ vs $T$ in Fig.$\refb{gvstgrand2}$ and it is clear that the global behavior is very similar to the canonical case as discussed in the paper. There are no HP transitions in this case.

For $N = 2/3$ case, the behavior of temperature $T$ and $C_{\phi}$ are different from the canonical case; $T >0$ only for small black holes and for those black holes, $C_{\Phi} <0$. Hence all physical black holes are locally unstable with negative specific heat capacity. We also noted that $G_{\Phi} <0$  for all positive $T$. In plotting $G_{\Phi}$ vs $T$, in Fig$\refb{gvstgrand1}$, one notice that there are no swallow tail type behavior leading to van der Waals type phase transitions or HP transitions. The preferred state is the black hole and not the radiation phase.

\begin{figure} [H]
\begin{center}
\includegraphics[scale=0.70]{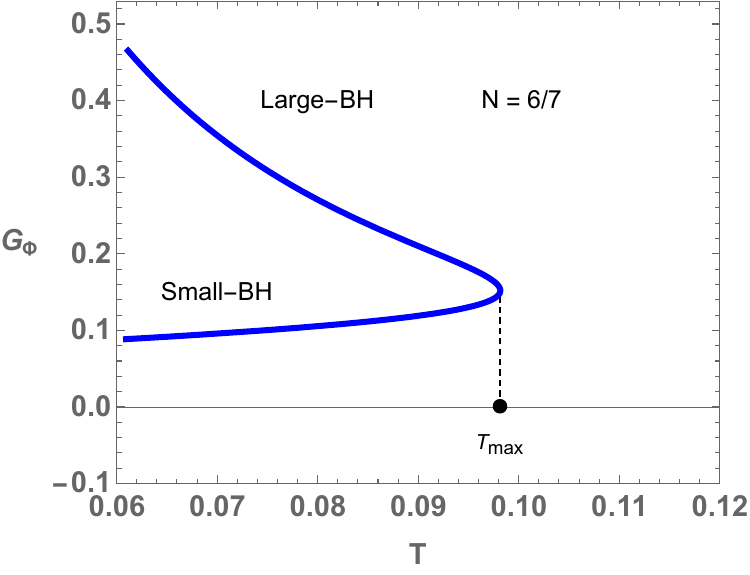}
\caption{$G_{\Phi}$ vs $T$ for $N = 6/7$ case for the grand canonical ensemble with fixed potential $\Phi$. Here, $\Lambda = -1$, and $\beta = 0.4, \gamma =1$, and  $\Phi = 0.6$.}
\label{gvstgrand2}
 \end{center}
 \end{figure}

\begin{figure} [H]
\begin{center}
\includegraphics[scale=0.70]{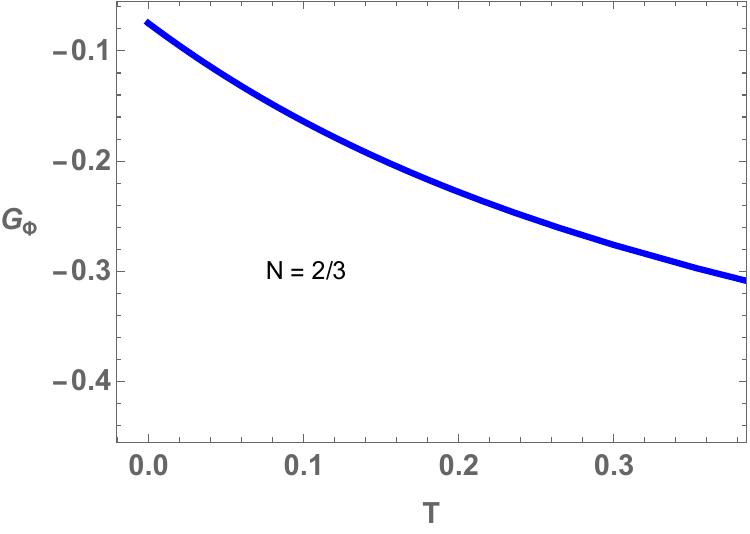}
\caption{$G_{\Phi}$ vs $T$ for $N = 2/3$ case for the grand canonical ensemble with fixed potential $\Phi$. Here, $\Lambda = -1$, and $\beta = 0.3, \gamma =1$, and  $\Phi = 0.5$.}
\label{gvstgrand1}
 \end{center}
 \end{figure}
In order to see if there are analogue of HP transition for other $N$ values, we studied $N=6/5$ case. For $N = 6/5$, we observed that both $T >0$ and $C_{\Phi} >0$ for the same values of $r_+$ as shown in Fig.$\refb{n65tandc}$. Hence the black hole are locally stable for positive temperature. $G_{\Phi}$ vs $T$ is plotted in Fig.$\refb{n65gvst}$. The behavior is very similar to the $G_{\Phi}$ vs T for charged BTZ black hole where there is a HP transition between large black hole and the radiation phase.
\begin{figure} [H]
\begin{center}
\includegraphics[scale=0.70]{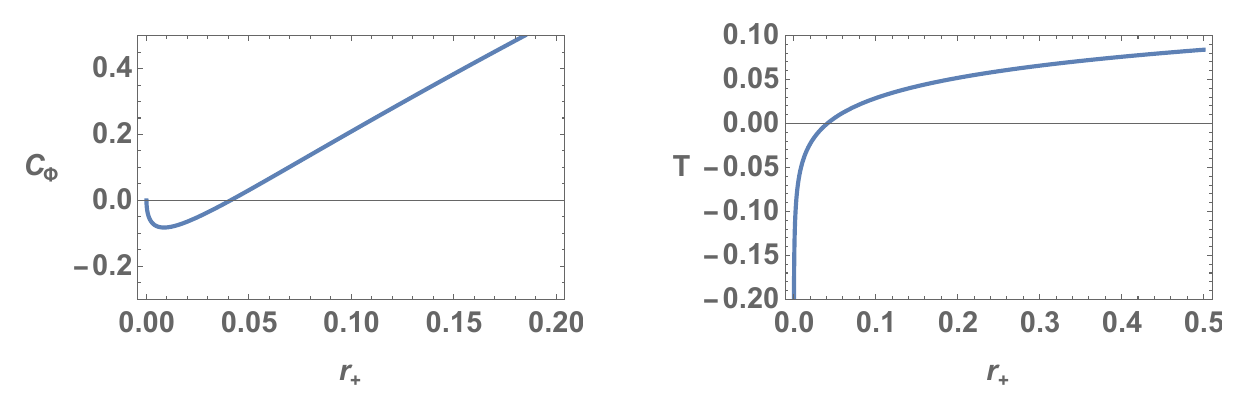}
\caption{The figure shows  $C_{\Phi}$ vs $r_+$ and $T$ vs $r_+$ for charged dilaton black hole with $N=6/5$. Here $\Phi = 0.5, \beta= 0.5, \gamma=1$ and  $\Lambda = -1$.}
\label{n65tandc}
 \end{center}
 \end{figure}
\begin{figure} [H]
\begin{center}
\includegraphics[scale=0.80]{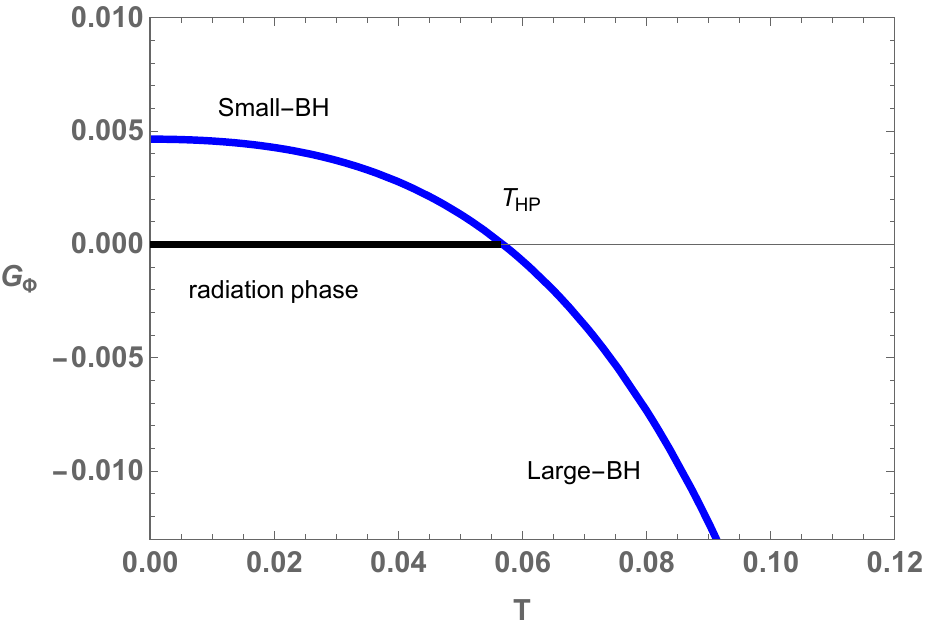}
\caption{The figure shows  $G_{\Phi}$ vs $T$  for charged dilaton black hole in grand canonical ensemble for $N=6/5$. Here $\Phi = 0.5, \beta=0.5, \gamma=1$ and  $\Lambda = -1$.}
\label{n65gvst}
 \end{center}
 \end{figure}

\subsection{Effect of the dilaton on Hawking-Page phase transition}
In summarizing the above studies from the grand canonical ensemble point of view, one can observe that the dilaton has a significant effect on the phase transitions. BTZ charged black hole do have HP transitions for all values of the parameters. For the charged dilaton black holes,  HP transition appear only for certain values of $N$. In our computations $N = 6/5$ has HP transition. On the other hand, $N = 1, 6/7, 2/3$ do not have HP transitions (or van der Waals phase transitions). Dilaton black holes also do not have re-entrant HP transitions. Hendi et al. \cite{hendi2} studied ensemble dependency of thermal stability of charged BTZ black holes. From our observation, charged BTZ black holes has similar thermodynamic behavior locally and globally both in canonical and grand canonical ensemble. For the dilaton black holes it is clear that the thermal stability depends on the ensemble (at least for some $N$ values) since there are clear differences of the thermodynamic behavior between the canonical ensemble and the grand canonical ensemble.


\section{Joule-Thomson expansion}

First, let us describe the Joule-Thomson expansion for a thermodynamical system as follows:
In a Joule-Thomson experiment, gas will flow constantly along a thermally insulated tube which is divided into two compartments. There would be a porous plug in between the compartments. The gas will be at higher pressure and will move into a low pressure section. The enthalpy will remain constant along the expansion process; hence it is a iso-enthalpic process. The purpose of this experiment is to find how the temperature $T$ varies with pressure $P$ at constant enthalpy $H$, which is given by $ \left(\frac{\partial T}{\partial P} \right)_H $ and is known as Joule-Thomson coefficient $\mu$.

We will omit the derivation of the Joule-Thomson coefficient since detailed derivations are given in \cite{okcu}. $\mu$ is given by,
\begin{equation} \label{mun}
\mu = \left(\frac{\partial T}{\partial P} \right)_H = \frac{1}{C_{P,Q}} \left[ T \left(\frac{\partial V}{\partial T} \right)_P - V \right]
\, .
\end{equation}
Here, $C_{P,Q}$ is the heat capacity at constant pressure. One can determine if the gas will cool or heat after the adiabatic expansion by evaluating the coefficient $\mu$. If $\mu >0 $ then the temperature decreases and if $\mu <0$ then the temperature increases. Note that in the Joule-Thomson expansion, change of pressure is negative. 

For the dilaton black hole considered in this paper, we will keep charge $Q$ constant when computing $\mu$. Hence $\mu$ is given by,
\begin{equation} \label{muwithr}
\mu =  \frac{1}{C_{P,Q}} \left[ T \left(\frac{\partial V}{\partial T} \right)_{P} - V \right] 
 = \frac{1}{C_{P,Q}} \left[ T \frac{\left(\frac{\partial V}{\partial r_+} \right)_{P}}{\left(\frac{\partial T}{\partial r_+} \right)_{P}}     - V \right] 
\end{equation}
\subsection{$\frac{2}{3} < N < 2 $ case}
In this section we will take thermodynamical volume V given in eq.$\refb{volumen}$, temperature T in eq.$\refb{tempn}$ and entropy S in eq.$\refb{entropy-S-N}$ to derive  $\mu$. Thermodynamical volume is given by,
\begin{equation} 
V =  \frac{4 \pi  \gamma \, \beta^{2 - N} r_+^{\frac{3 N}{2}-1}}{3 N - 2}  \, ,
\label{}
\end{equation}
Hence,
\begin{equation} \label{partialv}
\left(\frac{\partial V}{\partial r_+} \right) = 2 \pi \gamma \beta^{2 - N} r_+^{\frac{3 N}{2} - 2}
\end{equation}
Hawking temperature is given by,
\begin{equation}
 T  = \frac{ 8 \pi P \beta ^{2-N}  r_+^N - Q^2}{N \pi  r_+} 
  \, ,
\label{}
\end{equation}
Hence,
\begin{equation} \label{partialt}
\left(\frac{\partial T}{\partial r_+} \right) = \frac{ 8 \pi P (N-1) \beta^{2-N} r_+^N +  Q^2}{ \pi N  r_+^2}
\end{equation}
Entropy is given by,
\begin{equation}
S = \frac{2 \pi r_+ R(r_+)}{4} = \frac{\pi}{2} \gamma \, r_+^{N/2}    \, .
\label{}
\end{equation}
which can be applied to compute the specific heat capacity at constant pressure as,
\begin{equation}
C_{{P,Q}} =  T \left( \frac{\partial S}{\partial T} \right)_{P} =
T \frac{\left(\frac{\partial S}{\partial r_+} \right)_{P}}{\left(\frac{\partial T}{\partial r_+} \right)_{P}}
=
\frac{N \pi  \gamma \,   r^{N/2}_+ \left( \beta^2 \Lambda r^N_+  + Q^2 \beta ^N\right)}{(4 (N - 1)  \beta^2 \Lambda r^N_+ - 4 Q^2 \beta^N)} 
\label{C-N} 
\,.
\end{equation} 
By substituting eq.$\refb{partialv}$, eq.$\refb{partialt}$, $T$, $V$ and $C_{P,Q}$ into eq.$\refb{muwithr}$, one obtain the $\mu$ for the general values of $N$ as follows:
\begin{equation} \label{mu1}
 \mu = \frac{8 \beta^{2-N} r_{+}^{(N-1)}}{(3 N -2)}  \left( \frac{ - 3 Q^2 + 8 \pi  P \beta^{2-N} r_+^N }
{ - Q^2  + 8 \pi P \beta^{2-N} r_+^N} \right) \, .
\end{equation}
W would first like to make comments about $\mu$ in relation to horizon radius of the black hole. The divergence point of $\mu$ ($\mu \ra \infty$) occurs when the denominator of $\mu \ra 0$. This is the exact value of $r_+$ when $T =0$ leading to extreme black holes. Hence, divergence points correspond to the extremal black holes with $r_+ = r_{ex(N)}$   given by eq.$\refb{rex1}$. The zero of $\mu$ occurs when the  numerator of eq.$\refb{mu1}$ becomes zero at $r_+ = r_{\mu N}$ which is given by,
\begin{equation}    
r_{\mu N} = \left(\frac{ 3 Q^2}{ (-\Lambda \beta^{2 - N})} \right)^{1/N}
\end{equation}
It  is clear that $r_{\mu N} = 3 r_{ex(N)}$ which is larger than $r_{ex(N)}$. Hence, the inversion point, where $\mu =0$ does not correspond to a physical limit of the black hole.

In Fig.$\refb{muvsrn2}$, the Joule-Thomson coefficient $\mu$ is plotted against the horizon radius $r_+$ for different values of  $N$. One can observe that for large $N$, $\mu$ is higher. 
It is observed that for large $r_+$, $\mu$ reaches a constant value for all $N$ values. We have omitted the region where $\mu$ goes to $\infty$ since that is the point where the black hole becomes extreme; $r_+$ smaller than that will have no black holes. 
\begin{figure} [H]
\begin{center}
\includegraphics[scale=0.80]{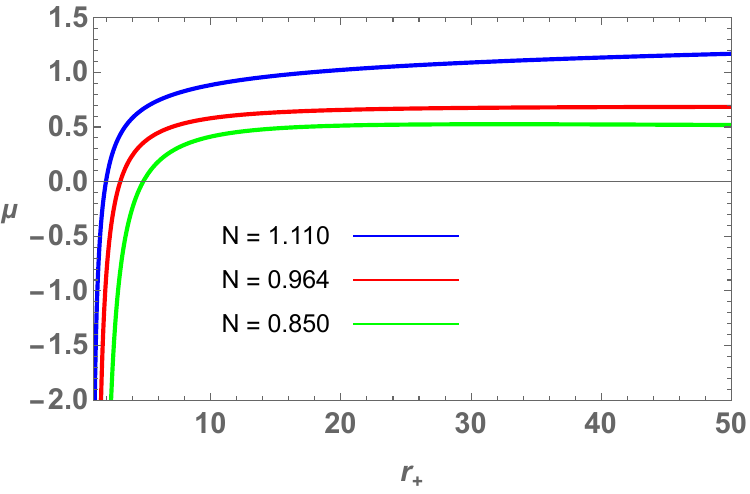}
\caption{The figure shows  the Joule-Thomson coefficient $\mu$  vs the horizon $r_+$ for various $N$ values. Here $Q = 1.5, P = 1, \beta = 0.1$.
}
\label{muvsrn2}
 \end{center}
 \end{figure}
In order to demonstrate how $T$ and $\mu$ are related, we have chosen $N=1$ as a representative for the N values in the range $2/3 < N < 2$ and plotted it in Fig.$\refb{mutemp}$. Here we have kept the charge Q fixed. It is clear that  $\mu$ and $T$ has similar behavior when plotted against $r_+$. Also, for large $r_+$, $\mu$ reaches a constant value for all $Q$.

\begin{figure} [H]
\begin{center}
\includegraphics[scale=0.85]{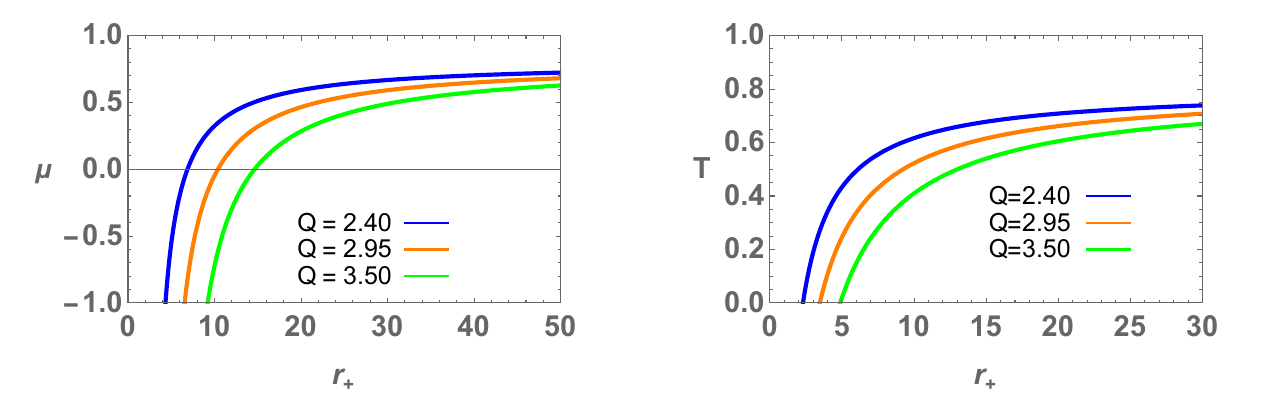}
\caption{The figure shows  the Joule-Thomson coefficient $\mu$ and Hawking temperature $T$ vs the horizon $r_+$ for $N =1$ case for fixed charge. Here $Q = 2.40, 2.95, 3.50$ and $P = 1, \beta = 0.1$.
}
\label{mutemp}
 \end{center}
 \end{figure}
One can define a quantity called  the inversion temperature, $T_i$ and inversion pressure, $P_i$ when $\mu =0$. $T_i$ is given by,
\be
T_i = V \left(\frac{\partial T}{\partial V} \right)_P \,  = \frac{ 2 Q^2}{ N \pi r_+}
\en
The gas will heat if one starts above the inversion temperature and will cool if one starts below the inversion temperature. The inversion pressure $P_i$ is calculated from $\mu=0$ via the eq.$\refb{mu1}$ as,
\be \label{pi}
P_i = \frac{ 3 Q^2}{ 8 \pi \beta^{(2-N)} r_+^N }  \, 
\en
By eliminating $r_+$ from both $P_i$ and $T_i$, one  can obtain a relation between the inversion temperature and inversion pressure as,
\be
T_i = \frac{ 2 Q^{ 2 - \frac{2}{N}}}{ N \pi} 
\left(\frac{ 8 \pi \beta^{ 2 - N}}{ 3} \right)^{1/N} P_i^{1/N}  \, 
\en
In Fig.$\refb{tivspin}$, $T_i$ vs $P_i$ is given for various $N$ values. It is clear that $T_i$ increases with $P_i$. Also, it is observed that there is only one value of $P_i$ for a given $T_i$ for all $N$ values. 
\begin{figure} [H]
\begin{center}
\includegraphics[scale=0.75]{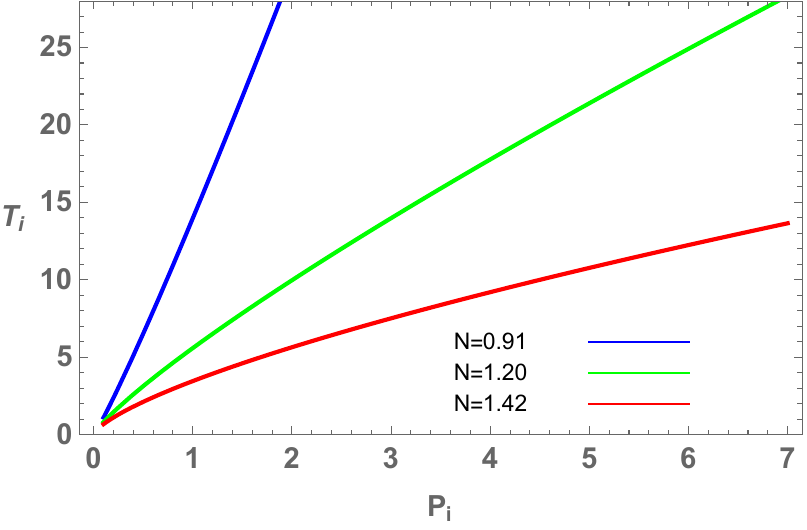}
\caption{The figure shows  the inversion curves, $T_i$ vs $P_i$   for various $N$ values.  Here $\beta = 1.88, Q = 1.636$.
}
\label{tivspin}
 \end{center}
 \end{figure}
For $ N =1$ case, the above relations simplifies to,
\be \label{ti}
T_i = \frac{ 2 Q^2}{ \pi r_+} = \frac{16 \beta} { 3} P_i
\en
It can be observed that for $ N =1$ the relation is independent of the charge $Q$. It only depends on $\beta$. The inversion temperature increases monotonically with the inversion pressure $P_i$ as shown in Fig.$\refb{tipi}$.

\begin{figure} [H]
\begin{center}
\includegraphics[scale=0.70]{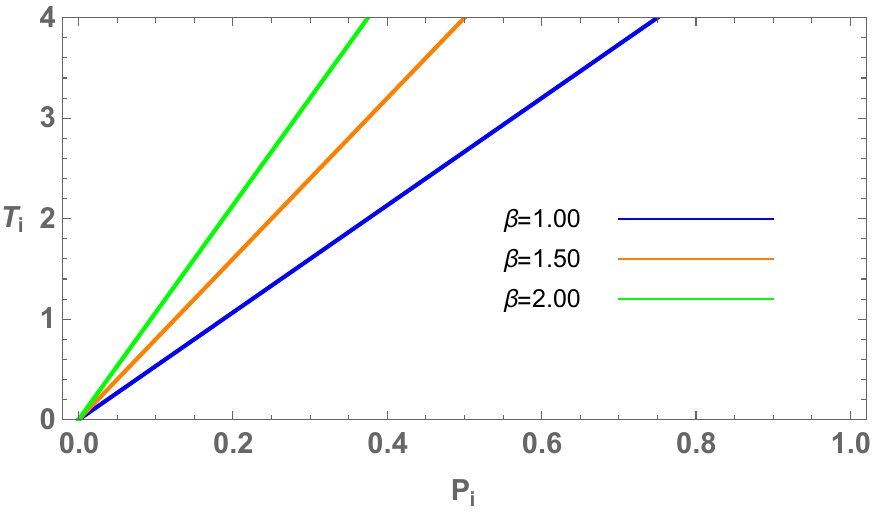}
\caption{The figure shows  the inversion curves, $T_i$ vs $P_i$  for the dilaton black hole for $N =1$ case.  Here $\beta = 1.00, 1.50, 2.00$.
}
\label{tipi}
 \end{center}
 \end{figure}
For a given $N$ and $Q$, there is only one inversion curve for the dilaton black hole. In contrast, for the van der Waals fluid, there are two inversion curves as shown in \cite{okcu}. Furthermore, the dilaton black hole does not have critical behavior similar to  van der Waals  fluids as shown in section (3.1). Hence a ratio between $T_i$ and $T_c$ does not exists. This behavior is similar to the BTZ black hole as shown in \cite{wang}.

Since the Joule-Thomson expansion is an isenthalpic process, we have plotted the isenthalpic curves for $N=1$  in Fig.$\refb{iso}$. Since in the extended phase space the mass is equal to the enthalpy (M = H), isenthalpic curves are the same as for constant $M$ curves. When $\mu=0$, i.e. $\left(\frac{\partial T}{\partial P} \right)_H =0$, the isenthalpic curves and inversion curves intersect at $\mu=0$ point. From Fig.$\refb{iso}$, it is clear that the slope of the isenthalpic curve is positive above the inversion curve. Hence, if the black hole starts above the inversion curve, it will cool under this expansion. On the other hand, since the slope of the isenthalpic curve is  is negative below the inversion curve, the temperature of the black hole will increase if it starts below the inversion curve.

\begin{figure} [H]
\begin{center}
\includegraphics[scale=0.70]{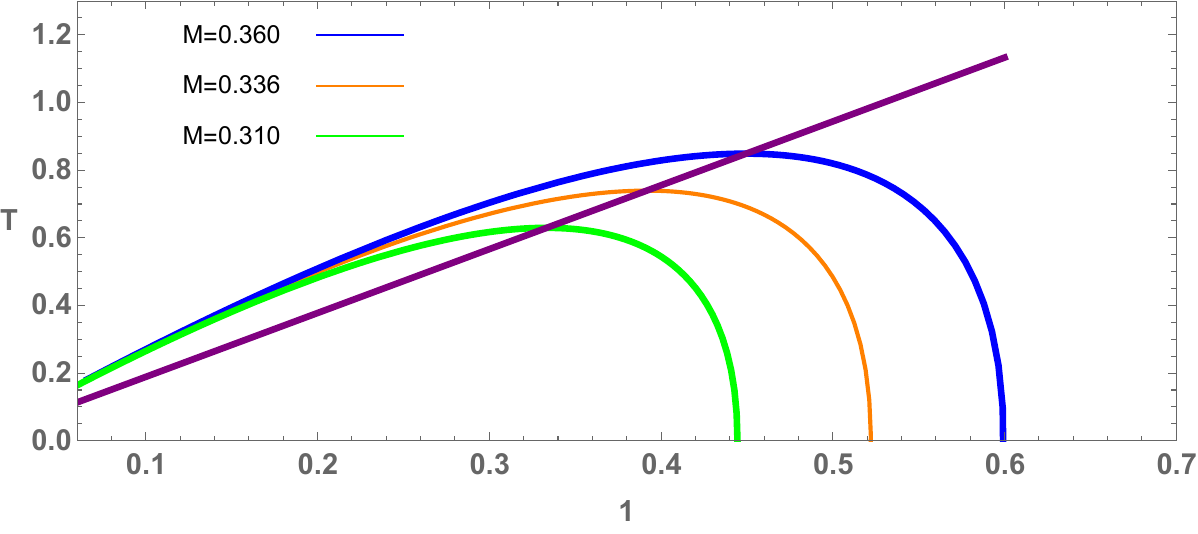}
\caption{The figure shows  isenthalpic and inversion curves for  $N =1$ case. The isenthalpic curve is given for mass M = H =0.310,0.336,0.360. The purple line is for inversion curve. Here, $Q = 1, \beta = 0.354$.
}
\label{iso}
 \end{center}
 \end{figure}

\subsection{$N = \frac{2}{3}$ case}

In this section we will take thermodynamical volume V given in eq.$\refb{volume23}$, temperature T in eq.$\refb{temp23}$ and entropy S in eq.$\refb{entropy23}$ to derive  $\mu$ for $N = \frac{2}{3}$. Thermodynamical volume is given by,
\begin{equation}
V =  2 \pi  \gamma \beta^{4/3} \log \left(\frac{r_+}{\beta }\right) \, .
\label{}
\end{equation}
Hence,
\begin{equation} \label{partialv2}
\left(\frac{\partial V}{\partial r_+} \right) = \frac{2 \pi \gamma \beta^{4/3}}{r_+}
\end{equation}
Hawking temperature is given by,
\begin{equation}
T =  -\frac{3 }{2 \pi r_+} \left( \beta ^{4/3} \Lambda  r_+^{2/3} +  Q^2 \right) \, 
\end{equation}
Hence,
\begin{equation} \label{partialt2}
\left(\frac{\partial T}{\partial r_+} \right) = \frac{ -8 \pi P  \beta^{4/3} r_+^{2/3} + 3 Q^2}{ \pi N \beta^N r_+^2}
\end{equation}
Entropy is given by,
\begin{equation}
S  = \frac{\pi}{2} \gamma \, r_+^{1/3}
\label{}
\end{equation}
which can be applied to compute the specific heat capacity at constant pressure as,
\begin{equation}
C_{P,Q} =   T \left( \frac{\partial S}{\partial T} \right)_{P} =
T \frac{\left(\frac{\partial S}{\partial r_+} \right)_{P}}{\left(\frac{\partial T}{\partial r_+} \right)_{P}}
= -\frac{\pi  \gamma r_+^{1/3}\left(Q^2  + \beta^{4/3} \Lambda  r_+^{2/3}\right)}{2( 3 Q^2 +  \beta^{4/3} \Lambda r_+^{2/3})}
\,\,\label{} \,
\end{equation}
By substituting eq.$\refb{partialv2}$, eq.$\refb{partialt2}$, $T$, $V$ and $C_{P,Q}$ into eq.$\refb{muwithr}$, one obtain the $\mu$ for $N = 2/3$ as follows:
\begin{equation}
\mu = \frac{12 \beta^{4/3}}{r^{1/3}} + \left(\frac{ 4 \beta^{4/3} ( 3 Q^2 - 8 \pi P r^{2/3} \beta^{4/3} ) }{r^{1/3}(Q^2 - 8 \pi P r^{2/3} \beta^{4/3})}\right) Log(r/\beta)
\end{equation}
The divergence point of $\mu$ where $\mu \ra \infty$ is when the denominator of $\mu$ goes to zero. In fact that is where the temperature also becomes zero leading to an extreme black hole at $r_+ = r_{ex}$ given by eq.$\refb{rex23}$. Due to the mathematical complexity, it is not possible to give an exact expression for $r_+$ when $\mu=0$. However, from the Fig.$\refb{mutemp2}$, where $\mu$ and $T$ are plotted against the $r_+$, one can observe that the horizon radius for $\mu=0$ is larger than the one for $T=0$ case. Hence the zero point of $\mu$ does not correspond to an extreme black hole where $T =0$. From Fig.$\refb{mutemp2}$, it is observed that the behavior of $\mu$ and $T$ are  very similar to the behavior for $2/3 < N < 2$ case. For large $r_+$, $\mu$ reaches a constant value for all $Q$.

\begin{figure} [H]
\begin{center}
\includegraphics[scale=0.85]{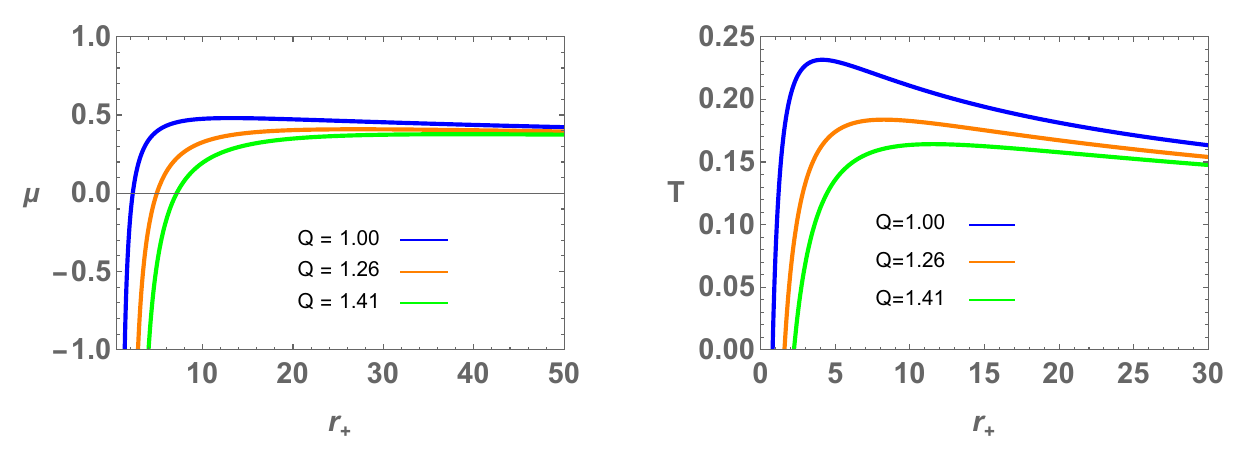}
\caption{The figure shows   $\mu$ and  $T$ vs  $r_+$ for $N =2/3$ case for fixed charge. Here $Q = 1.00,1.26,1.41$ and $P = 1, \beta = 0.1$.
}
\label{mutemp2}
 \end{center}
 \end{figure}
Inverse temperature $T_i$ and inverse pressure $P_i$ for $N=2/3$ case are given by,
\be
T_i = \frac{ 3 Q^2 ln(\frac{r_+}{\beta})}{ \pi r_+ ( 3 + ln(\frac{r_+}{\beta})) }
\en
\be
P_i = \frac{3 Q^2}{ 8 \pi r_+^{2/3} \beta^{4/3}} \left( \frac{ 1 + ln(\frac{r_+}{\beta})} {3 + ln(\frac{r_+}{\beta})} \right)
\en
Inversion curves for N=2/3 case is plotted in Fig.$\refb{tinvspin2}$. 
\begin{figure} [H]
\begin{center}
\includegraphics[scale=0.80]{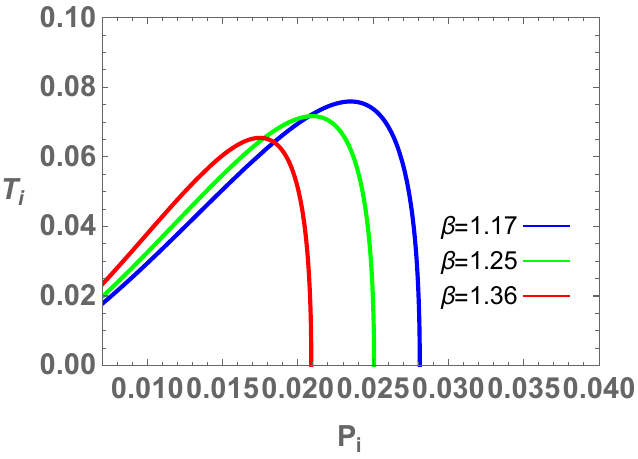}
\caption{The figure shows  the inversion curves, $T_i$ vs $P_i$   for $N =2/3$ case.  Here $Q=1$ and $\beta = 1.17, 1.25, 1.36$.
}
\label{tinvspin2}
 \end{center}
 \end{figure}
Isenthalpic curves for $N = 2/3$ is plotted in Fig.$\refb{ienthalpy2}$ and they are very similar to the $N=1$ case.
\begin{figure} [H]
\begin{center}
\includegraphics[scale=0.80]{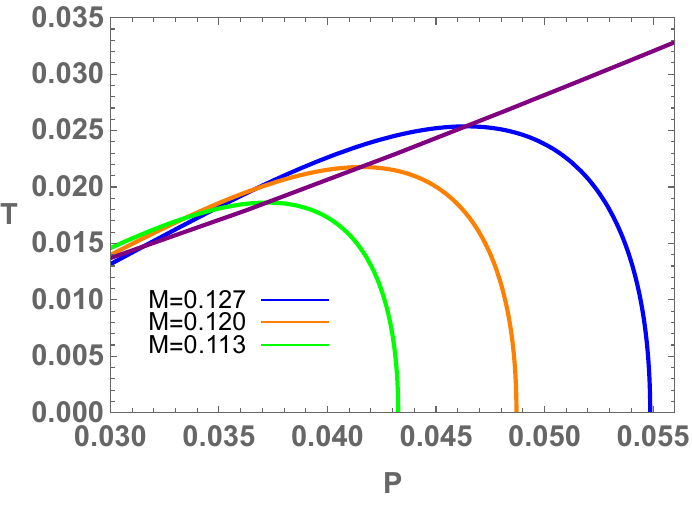}
\caption{The figure shows  isenthalpic and inversion curves for $N =2/3$ case. The isenthalpic curve is given for mass M = H = 0.127,0.120,0.113. The purple line is for inversion curve. Here, $Q = 0.395, \beta = 0.22$, and  $\gamma=1$.
}
\label{ienthalpy2}
 \end{center}
 \end{figure}

\subsection{Comparison of inversion curves for different thermodynamical volumes}
Thermodynamic volume differs for $2/3 < N <2$ case and N=2/3 case. Hence the inversion curves do get effected.  There is a significant difference in the behavior of $T_i$ vs $P_i$ when compared  the case for $N=1$ with $N=2/3$. For $N=1$ case, there is only one value of $P_i$ for a given $T_i$ where as  there are two possible values for $P_i$ for a given $T_i$. For both cases, when all the parameters are fixed, inversion point where  $\mu=0$ case yields  a horizon radius larger than the radius for the extreme black hole. Hence the inversion point does not correspond to any physical limit. Also from Fig.$\refb{iso}$ and Fig.$\refb{ienthalpy2}$,the inversion temperature, $T_i$ where the inversion curve intersects with the isenthalpic curve (inversion point) increases with mass $M$ for both $N=1$ and $N=2/3$ case. This suggests that the inversion temperature is higher for larger enthalpy for both cases.


\section{Reverse Isoperimetric Inequality}

There is a conjecture that the isoperimetric ratio $\cal{R}$ given by,
\be
{\cal{R}} = \left(\frac{ (D-1) V}{ \omega_{D-2}} \right)^\frac{1}{ D-1} \left( \frac{\omega_{D-2}}{ A}\right)^ \frac{1}{D-2}
\en
will always satisfy ${\cal{R}} \geq 1$ when $V$ is the thermodynamic volume, $A$ is the area of the horizon and $\omega_{D-2}$ is the area of the $(D-2)$-dimensional unit sphere \cite{cvetic}. The Reverse Isoperimetric Inequality  thus establishes a bound between the thermodynamic volume and the horizon area (or entropy) of black holes in the extended phase space. However, in the case of the charged BTZ black hole, we have ${\cal{R}} < 1$ (referred to as superentropic)~\cite{mann4}. A similar situation is observed in other solutions in 2+1 dimensions \cite{He:2017ujy, Ghosh:2020kba}. Nevertheless, from an alternative endpoint, it has been shown that when a new thermodynamic quantity, defined as a renormalization length scale is introduced, the inequality ${\cal{R}} \geq 1$ can be satisfied even in such cases.

On the other hand, several studies have examined the conjecture from a different perspective, associating it with the behavior of specific heat capacities. Ref. \cite{Johnson:2019mdp} conjectures that superentropic black holes always satisfy $C_V < 0$, implying thermodynamic instability in the extended phase space. However, this conjecture is refuted in Ref. \cite{Cong:2019bud} through a counterexample, and an alternative condition is proposed: any black hole that violates the Reverse Isoperimetric Inequality is thermodynamically unstable either because $C_V < 0$ or $C_P < 0$.

In the present work, $D = 3$ and $\omega_1 = 2 \pi$. For $2/3 < N < 2$, one finds
\be
A = 2 \pi  \gamma  r_+^{N/2}  \,\,  ; \hspace{0.5 cm} V = \frac{4 \pi  \gamma \, \beta^{2-N} r_+^{\frac{3 N}{2}-1}}{3N - 2}\,\, ; \hspace{0.5 cm} \omega_1 = 2 \pi 
\en
leading to
\be
{\cal{R}} = 2 \beta ^{1 - \frac{N}{2}} \sqrt{\frac{1}{(3 N - 2)\gamma}} r_+^{\frac{N-2}{4}} \, .
\en
On the other hand, for $N = 2/3$,
\be
A = 2 \pi  \gamma \, r_+^{1/3}\,\,  ; \hspace{0.5 cm} V = 2 \pi \gamma \, \beta ^{4/3}   \log \left(\frac{r_+}{\beta }\right)
\en
and
\be
{\cal{R}} = \frac{ \sqrt{2 \beta^{4/3} \log \left(\frac{r_+}{\beta }\right)}}{\gamma^{1/2}  \, r_+^{1/3}}\, .
\en

The parameter $\beta$ is fixed by physical boundary conditions, so modifying its value changes the asymptotic value of the dilaton field and consequently represents a genuine physical degree of freedom. Therefore, variations in $\beta$ change the horizon, and thus the entropy and the thermodynamic volume. As a result, the isoperimetric ratio ${\cal{R}}$  may or may not satisfy the reverse isoperimetric inequality depending on the chosen value of $\beta$.

In contrast, since $\gamma$ merely rescales the angular coordinate it does not describe a physical degree of freedom, but rather a normalization convention. Varying $\gamma$ does not change the horizon and therefore has no effect on the isoperimetric ratio ${\cal{R}}$. This is analogous to the renormalization length appearing in the logarithmic term of the charged BTZ solution \cite{Martinez:1999qi}, which is likewise not treated as a physical parameter and therefore does not affect the isoperimetric inequality.

Therefore, since $\beta$ is an integration constant and its value is not determined a priori, it is not possible to say whether ${\cal{R}} \geq 1$. Hence, inclusion of the dilaton field leads to the possibility of charged black holes in 2+1 dimensions to satisfy the Reverse Isoperimetric Inequality. Similarly, we can find parameter ranges where ${\cal{R}} < 1$.
As an illustration of both situations, Fig. $\refb{Reverse}$ shows the dependence of ${\cal{R}}$ on $\beta$, keeping the parameters $M$, $\Lambda$ and $q$ fixed.

\begin{figure} [H]
\begin{center}
\includegraphics[scale=0.80]{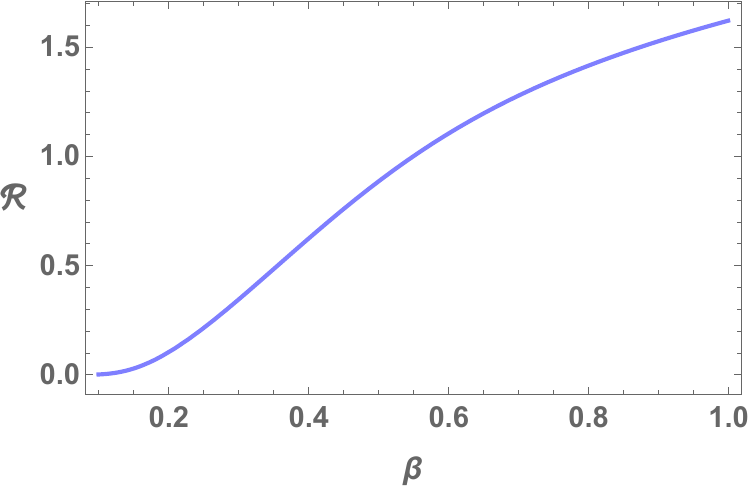}
\caption{Dependence of ${\cal{R}}$ on $\beta$, showing the regions where ${\cal{R}} < 1$ and ${\cal{R}} \geq 1$ for the case $N = 2/3$, with fixed parameters $M = 0.3$, $\Lambda = -1$ and $Q = 0.65$ ($\gamma = 1$).}
\label{Reverse}
 \end{center}
 \end{figure}

Similarly, we can find parameter ranges where ${\cal{R}} < 1$ as shown on the left in Fig.$\refb{Reverse CP}$.  In this case, there are values for the parameters for which $C_P > 0$ or $C_P < 0$ (with $C_V = 0$  in both cases), as illustrated on the right in Fig.$\refb{Reverse CP}$.

\begin{figure} [H]
\begin{center}
\includegraphics[scale=0.80]{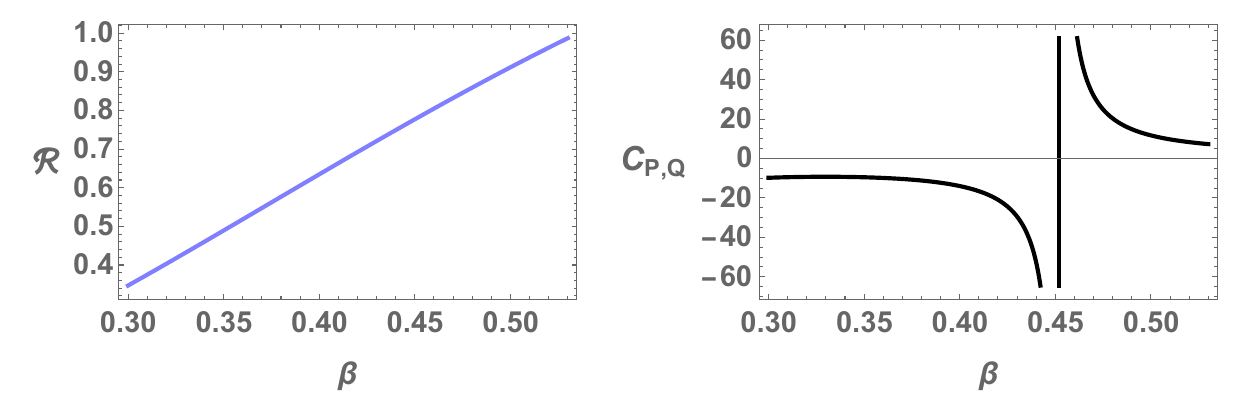}
\caption{Here we consider a case in which ${\cal{R}} < 1$ and $C_P$ exhibits both positive and negative values. We take $N = 2/3$, with fixed parameters $M = 0.3$, $\Lambda = -1$, $Q = 0.8$ and $\gamma = 1$.}
\label{Reverse CP}
 \end{center}
 \end{figure}


\section{Conclusion}
In this paper we have studied the static charged dilaton black hole in 2+1 dimensions
presented in Chan and Mann \cite{Chan:1994qa} and  Xu \cite{Xu:2019pap}. There are number of parameters of the black holes solution given by $N, M, Q, \gamma, \Lambda$ and  $\beta$.  Black holes were studied for various values of the parameter $N$: it is related to the coupling constant for the dilaton with the electromagnetic fields and the gravitational field. Parameter values for all the figures in the paper are given in the Appendix B in Table B.

Black hole solutions exist only for $2/3 \leq N < 2$. All the black holes for this range of $N$ values, has the possibility to have two horizons, degenerate  or no horizons depending on the mass and charge. We have studied thermodynamics in the extended phase space where $P = - \Lambda/ 8 \pi$. We have computed the Hawking temperature $T$ and the entropy $S$.  We have presented first law of thermodynamics for all the black holes considered in this paper. One has to consider $\beta$ as thermodynamic variable in order to make the first law of thermodynamics to be consistent. Other thermodynamical quantities such as specific heat, and Gibbs free energy are computed to analyze the local and global stability. State equation is also presented. When it comes to thermodynamic stability we observed two major groups: black holes with $2/3 \leq N < 1$ and black holes with $ 1 \leq N < 2$. Our main focus had been on  thermodynamics in the canonical ensemble, but we also have studied thermodynamics in the grand canonical ensemble in section 8.

For $2/3 \leq N < 1$ there is a maximum temperature  ($T_{max})$ at $r_+ = r_{max}$ and black holes cannot exist with temperature greater than $T_{max}$. Also, these black holes can have degenerate horizons at $r_+ = r_{ex}$ where the temperature $T =0$. The existence of a maximum temperature does effect the specific heat capacity $ C_{P,Q}$ significantly; there would be two branches for $C_{P,Q}$. $C_{P,Q}$ is positive for small black holes and $C_{P,Q}$ is negative for large black holes. $C_{P,Q} = 0 $ at $r_+ = r_{ex}$. Hence, locally, small black holes are stable. The range of $r_+$ for which the small black hols are stable is from  $r_{ex}$ to $r_{max}$. In order to understand the global stability, we have studied the Gibbs free energy $G$. To make quantitative observations we have focused on $N= 2/3$ and $N = 6/7$ cases. We observed that $G$ is smaller for the smaller black holes and is positive. Hence globally, small black holes are preferred. There are no first order  HP phase transitions, re-entrant HP phase transitions since the black holes are charged. There are no  van der Waals type phase transitions either.

For $ 1 \leq N < 2$, there is no maximum temperature. In order to study details of thermodynamic behavior, we focused on $N=1$ case. We observed that  for all black holes with  $T > 0$,  $C_{P,Q} >0$. Hence,  all black holes are locally stable. $G \geq 0$ for $N=1$ case and is continuous. Hence such black holes do not admit van der Waals type phase transitions. Globally, all black holes are stable:  there are no HP transitions or re-entrant HP phase transitions because black holes are charged.

In all three cases ($N = 1, 6/7$, and $2/3$), the heat capacity $C_{P,Q} = 0 $ whenever the temperature $T=0$ (at $r_+ = r_{ex}$). Physically, this indicates that the black hole reaches a state in which it cannot absorb heat to increase its temperature. At $T = 0$, the entropy no longer responds to changes in the thermodynamic variables, the near-horizon geometry becomes static and unalterable and the surface gravity vanishes, and the system cannot exchange heat with the environment. The heat capacity necessarily goes to zero, the black hole cannot increase its temperature through any thermodynamic process, consistent with the behavior expected for an extremal configuration.

In order to compare, we also presented the thermodynamics of neutral BTZ and the charged BTZ black holes. These are studies that are already done but for the sake of comparison we have presented them. We observed that the local and the global behavior of  BTZ (neutral)  black hole behave very similar to the N=1 dilaton black hole. The uncharged BTZ black hole is globally stable for all $r_+$ and differs from the N=2/3 and N=6/7 charged dilaton black holes. For charged BTZ black hole, larger black holes are preferred globally where as for $N=2/3$ and $N = 6/7$ smaller black holes are preferred. Hence the presence of the dilaton has made a difference in the thermodynamic behavior.

It was shown  that the uncharged $N = 1$ black hole is dual to the $N = 1$ black hole in string theory.  It would be interesting to see if there is an underlying relation between duality and thermodynamic stability.

Joule Thomson expansion is studied in detail and Joule Thomson coefficient $\mu$  is calculated. 
We observed that for large $N$, $\mu$ is higher. It is clear that $\mu$ diverges at $T =0$ where the black hole is extreme. Inversion temperature is calculated and is observed that it is proportional to the inversion pressure $P_i$ when $N=1$. For $N = 2/3$ the behavior is somewhat different. There could be two values of $P_i$ for a given $T_i$ in this case. Isenthalpic curves are plotted and we have discussed how the black hole heat or cool with respect to the inversion curve. It is interesting to observe that for $N=1$, the relation between the inversion temperature $T_i$ and the inversion pressure $P_i$ is independent of the charge $Q$. 

The Reverse Isoperimetric Inequality has been analyzed for charged black holes in 2+1 dimensions, including the presence of a dilaton field. Unlike other cases such as the charged BTZ black hole, the inclusion of the dilaton allows the inequality to be fulfilled for certain values of the integration parameter $\beta$. Parameter ranges where ${\cal{R}} < 1$ are also identified, showing that superentropic situations can occur even with dilaton. Thermodynamic stability reveals that in these cases $C_P$ transitions between positive and negative values while $C_V$ remains zero, indicating parameter-dependent instabilities. This shows that the relationship between the thermodynamic volume and the horizon area is sensitive to the values of the physical parameters and the inclusion of additional fields in the black hole geometry.

In this paper, the main focus has been on thermodynamics in the canonical ensemble where charge $Q$ was fixed. We also have studied the thermodynamical behavior in the grand canonical ensemble in section 8 to understand if there are HP transitions: here, we fix the potential $\Phi$. What we observed was quite interesting: while the charged BTZ black hole had a HP transitions, dilaton black hole had HP transitions only for $N = 6/5$. For all the other values of $N$ that we have studied in this paper ($N = 1, 2/3, 6/7$) HP transitions were absent. Also there were no re-entrant HP transitions for both charged BTZ black hole or the charged dilaton black hole. The dilaton black hole has a significant impact on the thermodynamics  of the charged black holes and their phase transitions.

In this paper we focused on the thermodynamics of the charged dilaton black hole in the canonical ensemble and the grand canonical ensemble. We plotted  $G$ vs $T$ graphs for uncharged versions of the black holes for $N =2/3, 6/7$ in Appendix A to understand if there are HP phase transitions for neutral dilaton black holes. They are in Fig.$\refb{gvstun23}$ and Fig$\refb{gvstun67}$. From observations of the figures, it is clear that $N=2/3$ case has HP transitions. It would be interesting to study other uncharged dilaton black holes in 2+1 dimensions to understand how the dilaton effects the HP phase transitions in 2+1 dimensions.

In extending this work, it would be interesting to study how adding spin to these black holes will change the thermodynamical  behavior. Hence, it would be interesting to study the spinning dilaton  black holes presented by Chan and Mann in \cite{Chan:1995wj}. For the canonical ensemble, the angular momentum $J$ would be fixed in this case.

\section*{Acknowledgments}

S. Fernando was on  Sabbatical leave from Northern Kentucky University when part of this work was completed.  L. Balart is supported by DIUFRO through the project: DI24-0087.


\section*{Data Availability Statement}
 No Data associated in the manuscript.

\appendix
\section*{Appendix}


\section{Thermodynamic stability for uncharged dilaton black holes}
\subsection{ $N= 2/3$ case}
Here, 
\begin{equation} \label{temp23}
T =  -\frac{3 \beta ^{4/3} \Lambda}  {2 \pi r_+^{1/3}} \, 
\label{}
\end{equation}
and
\begin{equation}
G = \frac{3 \gamma \beta^{4/3} \Lambda }{4} -\frac{1}{4} \gamma \beta^{4/3} \Lambda  \log \left(\frac{r_+}{\beta }\right)  
\,\,\label{} \,
\end{equation}
\begin{figure} [H]
\begin{center}
\includegraphics[scale=0.80]{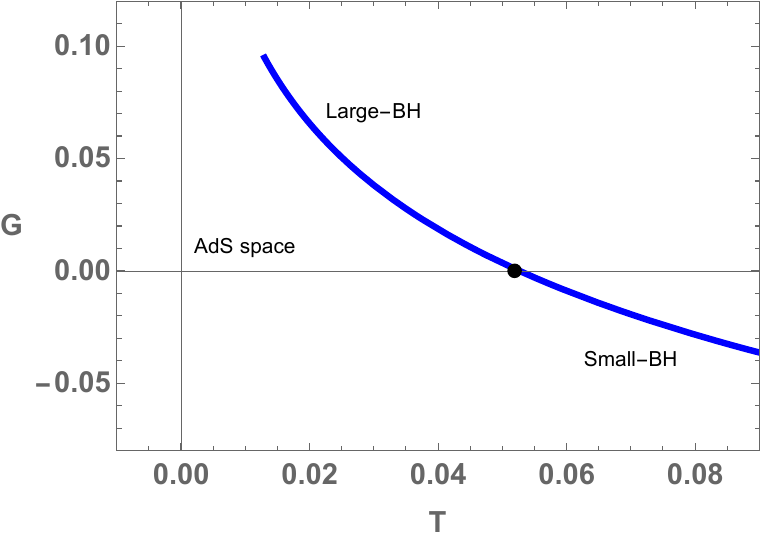}  
\caption{Gibbs free energy $G$ vs $T$ for uncharged black hole with  $N = \frac{2}{3}$. Here, $\beta = 0.3, \gamma = 0.45, \Lambda = -1$.}
\label{gvstun23}
 \end{center}
 \end{figure}
When $T$ is gradually increased, there is a HP phase transition between the thermal-AdS space and small black hole.
\subsection{ $N = 6/7$ case}
\begin{equation}
T =  \frac{-7 \beta^{8/7} \Lambda}{ 6 \pi r_+^{1/7}}    \, .
\label{temp2}
\end{equation}
\begin{equation}
G  =  \frac{-7 \beta^{8/7} r_+^{2/7}\Lambda}{ 24}
\label{temp2}
\end{equation}
\begin{figure} [H]
\begin{center}
\includegraphics[scale=0.80]{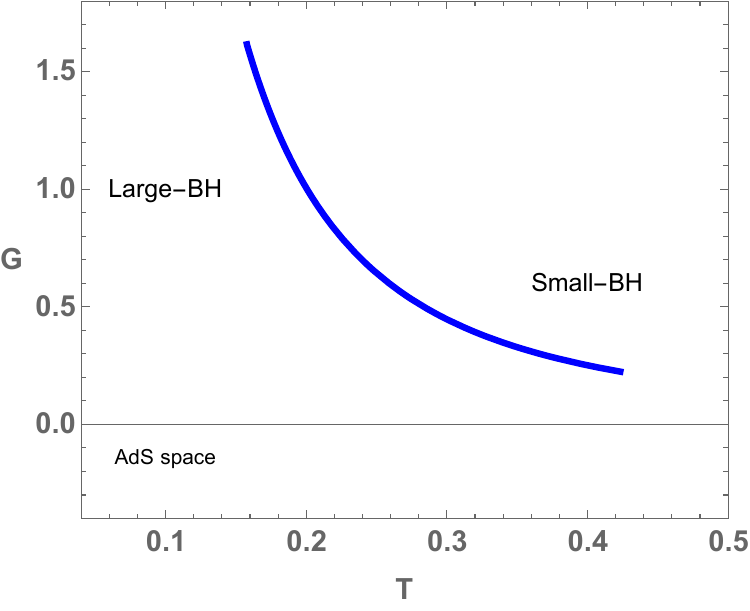}  
\caption{Gibbs free energy $G$ vs $T$, for uncharged black hole with  $N = \frac{6}{7}$. Here, $\beta = 1, \gamma = 1, \Lambda = -1$.}
\label{gvstun67}
 \end{center}
 \end{figure}
When $T$ is gradually increased $G$ decreases but always remains positive. Thermal-AdS space is preferred state since it has lower $G$. Black holes are locally stable (since $C_P = 3 \pi r_+^{3/7}/2 >0 $) but globally unstable. There are no HP transitions.}

\newpage

\section{ Table for all the parameters  in the figures in the paper}

\begin{center}
{\begin{tabu}{|l|l|l|l|l|l|l|l|l|r} \hline \hline 
\rowfont{\color{black}}
Figure  &  $\gamma$  &  $\beta$ & Q & $\Lambda$ & N & M & T & $\Phi$\\ \hline
\rowfont{\color{black}}
Fig.$\refb{tvsn}$ &   -  & 12.88  & 1 & -0.562 & 1.14,1.02,0.88,0.82 & - & - & - \\ \hline
\rowfont{\color{black}}
Fig.$\refb{cpalln}$ &   1  & 17(L),0.5(R)  & 0.5(L), 0.386(R) & -1 & 0.84, 0.89, 0.92(L) & - & - & -\\ 
\rowfont{\color{black}}
        &          &              &   &      & 1.52,1.36,1.13(R)  & - & - & -\\ \hline
\rowfont{\color{black}}
Fig.$\refb{frr}$ &   1  & 0.1  & 0.37 & -1 & 1, 6/7 & 0.14,0.12,0.10 & - & -\\ \hline
\rowfont{\color{black}}
Fig.$\refb{pvsrn}$ &   -  & 0.1  & 0.259 & - & 6/7 & - & 0.406 & - \\ \hline
\rowfont{\color{black}}
Fig.$\refb{cptn1}$ &   1  & 0.97,0.64,0.49  & 0.5 & - 1 & 6/7 & - & - & -\\ \hline
\rowfont{\color{black}}
Fig.$\refb{cptn2}$ &   1  & 0.97,0.64,0.49  & 0.5 & - 1 & 6/7 & - & - & -\\ \hline
\rowfont{\color{black}}
Fig.$\refb{gvstn}$ & 1 & 1 & 0.5 & -1 & 6/7 & - & - & -\\ \hline
\rowfont{\color{black}}
Fig.$\refb{gandtn}$ & 1 & 1 & 0.5 & -1 & 6/7 & - & - & -\\ \hline
\rowfont{\color{black}}

Fig.$\refb{cpt}$ & 1 & 0.5,1,2 & 0.5 & -0.2 & 1 & - & - & -\\ \hline
\rowfont{\color{black}}
Fig.$\refb{GvsT1}$ & 1 & 0.5,1,2 & 0.5 & -0.2 & 1 & - & - & -\\ \hline
\rowfont{\color{black}}
Fig.$\refb{f-23}$ & 1 & 0.7  & 0.95,0.65,0.81 & -1 & 2/3 & 0.3 & - & - \\ \hline
\rowfont{\color{black}}
Fig.$\refb{pvst23}$ & - & 1 & 1 & - & 2/3 & - & 0.5 & - \\ \hline
\rowfont{\color{black}}
Fig.$\refb{gvst3}$ & 1 & 0.48,0.40,0.36 & 0.5 & -1 & 2/3 & - & - & - \\ \hline
\rowfont{\color{black}}
Fig.$\refb{gandt3}$ & 1 & 0.48,0.40,0.36 & 0.5 & -1 & 2/3 & - & - & -\\ \hline
\rowfont{\color{black}}
Fig.$\refb{gvstcbtz}$ & - & - & 0.5 & - 1 & - & - & - & -\\ \hline
\rowfont{\color{black}}
Fig.$\refb{cptgrandBTZ}$ & - & - & - & - 1 & - & - & - & 0.5\\ \hline
\rowfont{\color{black}}
Fig.$\refb{gvstgrandBTZ}$ & - & - & - & - 1 & - & - & - & 0.5\\ \hline
\rowfont{\color{black}}
Fig.$\refb{gvstgrand2}$ & 1 & 0.4  & - &  - 1 & 6/7 & -  & - & 0.6 \\ \hline 
\rowfont{\color{black}}
Fig.$\refb{gvstgrand1}$ & 1 & 0.3  & - &  - 1 & 2/3 & -  & - & 0.5\\ \hline
\rowfont{\color{black}}
Fig.$\refb{n65tandc}$ & 1 & 0.5  & - &  - 1 & 6/5 & -  & - & 0.5\\ \hline
\rowfont{\color{black}}
Fig.$\refb{n65gvst}$ & 1 & 0.5  & - &  - 1 & 6/5 & -  & - & 0.5\\ \hline

\rowfont{\color{black}}
Fig.$\refb{muvsrn2}$ & - & 0.1 & 0.5 & - 25.13 & - & - & - & -\\ \hline
\rowfont{\color{black}}
Fig.$\refb{mutemp}$ & - & 0.1 & 2.40,2.95,3.50 & - 25.13 & 1 & - & - & -\\ \hline
\rowfont{\color{black}}
Fig.$\refb{tivspin}$ & - & 1.88 & 1.636 & -  & - & - & - & -\\ \hline
\rowfont{\color{black}}
Fig.$\refb{tipi}$ & - & 1.0,1.5,2.0 & - & -  & 1 & - & - & -\\ \hline
\rowfont{\color{black}}
Fig.$\refb{iso}$ & - & 0.354 & 1 & -  & 1 & 0.310,0.336,0.360 & - & -\\ \hline
\rowfont{\color{black}}
Fig.$\refb{mutemp2}$ & - & 0.1 & 1.00,1.26,1.41 & -25.13  & 2/3 & -& - & -\\ \hline

\rowfont{\color{black}}
Fig.$\refb{tinvspin2}$ & 1 & 1.17,1.25,1.36 & 1 &   & 2/3 & - & - & -\\ \hline
\rowfont{\color{black}}
Fig.$\refb{ienthalpy2}$ & 1 & 0.22 & 0.395 &  - & 2/3 & 0.127,0.120,0.113& - & -\\ \hline

\rowfont{\color{black}}
Fig.$\refb{Reverse}$ & 1 & -  & 0.65 &  - 1& 2/3 & 0.3 & - & -\\ \hline
\rowfont{\color{black}}
Fig.$\refb{Reverse CP}$ & 1 & -  & 0.8 &  - 1& 2/3 & 0.3 & - & -\\ \hline
\rowfont{\color{black}}
Fig.$\refb{gvstun23}$ & 0.45 & 0.3  & 0 &  - 1 & 2/3 & -  & - & - \\ \hline 

\rowfont{\color{black}}
Fig.$\refb{gvstun67}$ & 1 & 1  & 0 &  - 1 & 6/7 & -  & - & - \\ \hline \hline

\end{tabu} }
\label{tab1}
\vspace{0.3 cm}

{Table B: Data for all figures in the paper}
\end{center}


\end{document}